\documentclass[a4paper,11pt]{article}
\pdfoutput=1

\usepackage{jcappub}
\usepackage[T1]{fontenc}
\usepackage[utf8]{inputenc}
\usepackage{amsmath,amssymb}
\usepackage{booktabs}
\usepackage{graphicx}
\usepackage{float}
\usepackage{xcolor}
\usepackage{url}

\newcommand{\Rgeo}{R_{\rm geo}}
\newcommand{\Rmez}{R_{\rm mean\text{-}z}}
\newcommand{\Rslope}{R_{\rm slope}}
\newcommand{\Rdirect}{R_{\rm direct}}
\newcommand{\Rjoint}{R_{\rm joint}}
\newcommand{\Om}{\Omega_m}

\newcommand{\dd}{\mathrm{d}}
\newcommand{\healpix}{HEALPix}
\newcommand{\ap}[1]{\left(#1\right)}
\newcommand{\bs}[1]{\left[#1\right]}
\newcommand{\arcmin}{\ensuremath{^{\prime}}}

\title{Field-level zero-lag statistics of the weak-lensing ratio:
methodology and systematics in a CSST-like survey}

\author[a,b]{Ji Yao}
\author[a]{Yuan Shi}
\author[c]{Chengliang Wei}
\author[a]{Lin Nie}
\author[a,b]{Huanyuan Shan}
\author[d,e,f]{Pengjie Zhang}
\affiliation[a]{Shanghai Astronomical Observatory (SHAO), CAS, Shanghai 200030, China}
\affiliation[b]{University of Chinese Academy of Sciences, Beijing, 101408, China}
\affiliation[c]{Purple Mountain Observatory, Nanjing 210008, China}
\affiliation[d]{Department of Astronomy, Shanghai Jiao Tong University, Shanghai 200240, China}
\affiliation[e]{Key Laboratory for Particle Astrophysics and Cosmology (MOE), Shanghai 200240, China}
\affiliation[f]{Tsung-Dao Lee Institute, Shanghai Jiao Tong University, Shanghai 200240, China}
\emailAdd{ji.yao@shao.ac.cn}
\emailAdd{hyshan@shao.ac.cn}

\abstract{\small 
Weak gravitational lensing is one of the most important cosmological probes of the coming decade, but the statistical information that can be extracted from the shear field is still far from exhausted. Field-level inference is among the most promising routes beyond two-point statistics, yet a systematic field-level study of the lensing ratio---the ratio of the lensing signals at two source redshifts, a geometric probe that is independent of the large-scale structure and galaxy bias---is still missing. We develop a field-level lensing ratio estimator based on zero-lag pixel statistics, and compare it with the conventional shear two-point (2pt) ratio on a CSST-like mock catalog built for the early data release of a Stage-IV survey. We investigate how different ratio estimators and observational systematics affect the ratio measurements and the resulting cosmological constraints.

Our findings are as follows: 
[1] The four ratio estimators agree in reasonable range for each individual triple, and they generally follows the $\Lambda$CDM redshift geometry, but lies about $1.5\%$ below the $n(z)$-weighted theoretical prediction, highlighting the importance of accurate ratio modeling. 
[2] In both the pure Gaussian shape noise case and the case including intrinsic alignments (IA, with non-Gaussian intrinsic shape), the fiducial field-level direct ratio estimator is more powerful than the two-point ratio, while in terms of $\Om$ it improves the constraining power by roughly $10\%$ (Gaussian noise) to $40\%$ (with IA) relative to the two-point ratio. Additionally, the combination of the two methods can further improve the cosmological output slightly. 
[3] Lens magnification is the largest contamination of the field-level lensing ratio and biases the resulting cosmology significantly. 
[4] We show that the field-level slope ratio estimator has a larger bias in the derived $\Om$, which results from a problematic covariance, and the direct ratio estimator is more robust. 
Overall, we see even without machine learning, the zero-lag statistics based field-level lensing ratio can play a more important role in improving the $\Omega_m$ constraint and benefit both large-scale structure and expansion-history studies.
}


\hypersetup{
  pdftitle={Field-level measurement of the weak-lensing ratio: methodology and systematics in a CSST-like survey},
  pdfauthor={Ji Yao, Yuan Shi, Chengliang Wei, Lin Nie, Huanyuan Shan, Pengjie Zhang},
  pdfkeywords={weak gravitational lensing, galaxy-galaxy lensing, cosmic magnification, dark energy, galaxy surveys}
}

\compress

\begin{document}
\maketitle
\flushbottom

% ===========================================================================
\section{Introduction}
\label{sec:intro}
% ===========================================================================

Weak gravitational lensing is one of the most important cosmological probes of the coming decade.  By mapping the coherent distortion of distant galaxy images, it traces the matter distribution and the geometry of the Universe in
a way that is independent of the details of galaxy formation
\cite{bartelmann2001,kilbinger2015}.  The current Stage-III surveys \cite{desy3,kids1000,hscy1,wright2025,anbajagane2025} have established cosmic shear as a precision probe of the late-time expansion history and the
growth of structure, and the next generation, including Euclid
\cite{euclid2020}, Vera Rubin LSST \cite{lsstdesc2018}, the Nancy Grace Roman Space Telescope \cite{cao2026roman}, and the Chinese Space Station Telescope (CSST) \cite{zhan2011,gong2019}, will improve the lensing signal-to-noise (S/N) by an order of magnitude. Most of
the cosmological information extracted so far comes from two-point statistics with well-understood statistical and systematic errors \cite{chisari2025,newman2022,li2023skils,liu2026baryon,yao2023kids1000}, but they discard the spatial
information beyond the two-point level. Because the underlying density
field and the resulting lensing field are highly non-Gaussian, a significant part of its cosmological information resides beyond two-point statistics \cite{porqueres2023,zhou2024,ding2025shear2cosmo} and is waiting for exploration.

Field-level inference, or sometimes referred to as simulation-based inference, is considered to be one of the most promising ways to go beyond two-point statistics. Instead of compressing the data into a correlation function, it normally uses machine learning to extract the information into a latent space and compares them with the simulations. The approach has been developed for cosmic shear and for the matter distribution \cite{porqueres2023,zhou2024,chen2026cluster,scognamiglio2026} and other high-order statistics, i.e., void lensing \cite{xu2026neutrino,su2025void}. The lensing \emph{ratio}---the ratio of the lensing signals measured for two source populations cast by the same lenses---\cite{schneider2016,schneider1992,schneider2002,sanchez2021shear,jaintaylor2003,bernsteinjain2004,zhang2005} is a particularly clean target for a field-level analysis. We expect a local (zero-lag) field-level ratio to extract more information from the field while remaining immune to the modeling complexity of the matter clustering \cite{mead2020,zhou2023halo,chen2025emulator2,chen2025emulator1}, of the galaxy bias of the lens sample \cite{mandelbaum2006,prat2022,zhou2025emulator3,wang2019clustering}, and of the baryonic feedback that reshapes the matter distribution \cite{zheng2024feedback,zhang2026feedback,zhou2026dm}. The ratio therefore isolates the geometry of the Universe while
being insensitive to the large-scale structure.  Despite these advantages, the field-level implementation of the lensing ratio has not been studied systematically.

In this paper, instead of using machine learning to summarize the field as a whole, we examine the field-level information using simple zero-lag pixel statistics with an appropriate Gaussian kernel. We develop a field-level lensing ratio pipeline and test it on a CSST-like mock catalog built for the early data release of a Stage-IV survey.  The mock is based on the Jiutian-1G simulation \cite{han2025} and contains seven tomographic bins, enough to form all 35 lens--source1--source2 triples.  We implement four estimators: the shear two-point ratio, the slope ratio estimated from the convergence-density relation, the direct ratio of Sun et al.\ \cite{sun2023}, which marginalizes over a common template, and a joint field+two-point estimator.  We apply them to the fiducial triple and to the full 35-triple scan, and we compare their constraining power and their response to intrinsic alignments, cosmic magnification, source-side shear biases, and photo-$z$ uncertainties.

The paper is organized as follows.  Section~\ref{sec:data} describes the Jiutian-based mock and the systematics levels. Section~\ref{sec:methods}
defines the estimators and the flat $\Lambda$CDM inference.
Section~\ref{sec:fiducial} presents the lensing ratio measurements for a fiducial
triple. Section~\ref{sec:scan} presents the 35-triple scan of all seven
tomographic bins, the systematics ladder, and the flat $\Lambda$CDM constraints. Section~\ref{sec:discussion}
discusses the performance of the field-level estimator, its sensitivity to systematics, and future applications. The appendices contain the smoothing-scale stability of the direct ratio, the covariance structure, error diagnosis on $\Om$, and the residual shear and mean-$z$ bias model.

% ===========================================================================
\section{Simulation and the mock catalog}
\label{sec:data}
% ===========================================================================

\subsection{The Jiutian-1G simulation}
\label{sec:jiutian}

The mock is built from the Jiutian-1G simulation, one of the high-resolution primary runs of the Jiutian suite designed for the CSST extragalactic surveys \cite{han2025}.  Jiutian-1G simulates $6144^3$ dark-matter particles in a $1000\,h^{-1}{\rm Mpc}$ comoving box, with a force softening of $4\,h^{-1}{\rm kpc}$ and a particle mass of
$3.7\times10^{8}\,h^{-1}M_\odot$, and adopts the Planck 2018 cosmology ($\Om=0.3111$, $\Omega_b=0.049$, $h=0.6766$, $n_s=0.9665$, and $\sigma_8=0.8102$) \cite{planck2018}. Galaxies are populated with the GAEA semi-analytical model applied to HBT+ merger trees; their spectral energy distributions and broadband magnitudes
are computed with StarDuster, and light-cone catalogs up to $z=5$ are constructed with BLiC \cite{tan2025}. 

From the resulting catalog, we use galaxy positions (RA, Dec) (including lens positions and unlensed positions), stellar masses $M_*$, redshifts $z$, true shear $(\gamma_1,\gamma_2)$ and convergence $\kappa$, the intrinsic-alignment field \cite{hirata2004,bridle2007} ($e^{I}_1,~e^{I}_2$), and photometric redshift estimates $z_p$ (the scatter of $z_p$ around the
true redshift, $\sigma_{z_p}\simeq0.05(1+z)$ \cite{hoyle2018}).

\subsection{The CSST-like mock}
\label{sec:mock}

\begin{figure}[tbp]
\centering
\includegraphics[width=\textwidth]{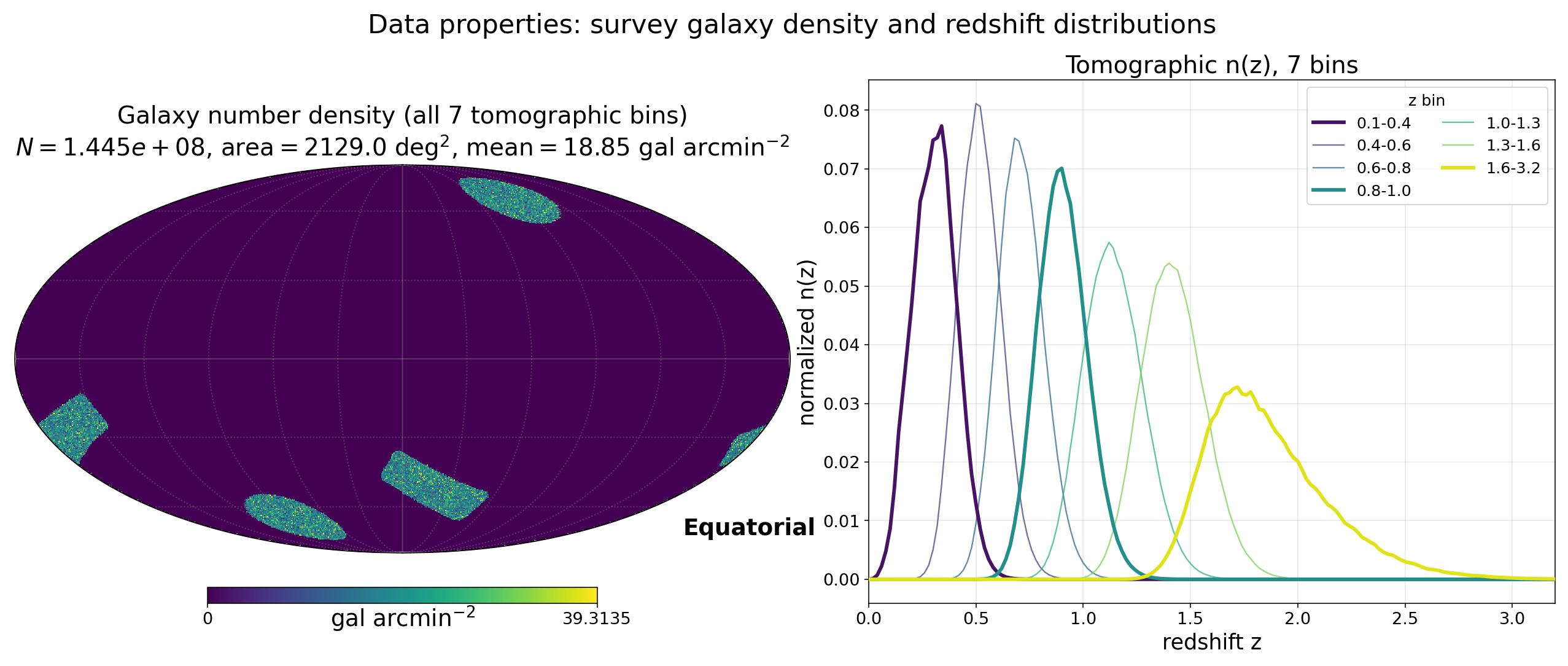}
\caption{Properties of the CSST-like mock catalog.  Left: \healpix\ map at $\rm nside=4096$ of all galaxies in the seven tomographic bins, in equatorial coordinates; the color scale is the galaxy number density. Right: normalized true-redshift distribution $n(z)$ of each tomographic bin. The three bins of the fiducial triple (photo-z range $0.1$--$0.4$, $0.8$--$1.0$, and $1.6$--$3.2$) are drawn with thick lines.}
\label{fig:data}
\end{figure}

To mimic a Stage-IV-like survey, we select
the sample by stellar mass $M_*$ and luminosity distance $D_L$ that satisfy $\log_{10}(M_*^{3/2}/D_L^2)>6.6$. This cut produces $N=1.445\times10^{8}$ galaxies over an effective area of $2129\ \rm deg^2$, corresponding to a mean surface density of $18.85\ \rm arcmin^{-2}$. The area is measured from the occupied pixels of the \healpix\ maps \cite{gorski2005} at $\rm nside=4096$ \footnote{The particles of the simulation are stored at $\rm nside=8192$.}. The galaxies are split by $z_p$ into the seven disjoint tomographic bins $0.1$--$0.4$, $0.4$--$0.6$, $0.6$--$0.8$, $0.8$--$1.0$, $1.0$--$1.3$, $1.3$--$1.6$, and $1.6$--$3.2$. Such tomographic binning makes the number of galaxies in each redshift bin reasonably similar. Figure~\ref{fig:data} shows the sky distribution of the sample and the
normalized $n(z)$ of each bin.

The seven redshift bins give 35 lens--source1--source2 triples with
$z_l<z_{s,1}<z_{s,2}$ for the lensing ratio measurements. From these, we select a fiducial triple as a detailed demonstration, with lens bin $0.1$--$0.4$, and the two source bins
$0.8$--$1.0$ and $1.6$--$3.2$. We denote this triple as
$0.1$--$0.4\,|\,0.8$--$1.0\,|\,1.6$--$3.2$, and later other triples will be presented in similar forms. It has a large geometric lever arm (the angular-diameter distances to the three bins change significantly) and a low lens redshift, which keeps the lens-magnification bias small. The lens bin contains $N_l=1.415\times10^{7}$ galaxies, and the two source bins contain $N_{s1}=2.132\times10^{7}$ and $N_{s2}=2.127\times10^{7}$ galaxies, with mean true redshifts $\bar z_l=0.308$, $\bar z_{s1}=0.905$, and
$\bar z_{s2}=1.864$. 

The mock is a single realization. We use the jackknife resampling errors to describe the survey-volume uncertainty within this realization, so that the realization-to-realization scatter (i.e., the super-sample covariance) is not sampled. A multi-realization study is left to future work.

\subsection{Observational systematics}
\label{sec:systematics}

We use the above mock catalog to build six data sets to explore the impact from different systematics, with each data set making one additional change from the previous: \\
(1) Gaussian noise only (with rms shape noise $e_{\rm rms}=0.275$), no intrinsic alignment, unlensed lens positions, true-$z$ $n(z)$. \\
(2) Add intrinsic alignment (IA) \cite{wei2018ia}: central galaxies follow the shape of their host halo (i.e., tidal alignment) \cite{hirata2004,bridle2007,yao2020ia}, while the satellite galaxies follow the spin of the halo (i.e., tidal torquing) \cite{blazek2015,blazek2019,mccullough2026}. A Gaussian scatter of $\sigma_e=0.2$ is added to the shape, i.e., $e=\gamma+e^I+\mathcal{N}(0,\sigma_e^2)$, so that the net shape noise is maintained at $e_{\rm rms}=0.275$. We note that $\gamma$ and $e^I$ are highly non-Gaussian fields, and $e^I$ contributes significantly to the net rms error, while the contribution from $\gamma$ is negligible. \\
(3) Adding lens magnification \cite{elvinpoole2023,lange2024,yao2023kids1000,qin2024,qin2025}: on top of data set (2), the lens catalog uses the lensed positions, so that the cosmic magnification is included. The lensed galaxy number density $\delta_g^L=\delta_g+g_{\rm mag}\kappa_g$ will contain not only the intrinsic clustering $\delta_g$, but also the lensing magnification, with the prefactor $g_{\rm mag}=2(\alpha-1)$, and $\alpha=0$ in this simulation as we only include the area-change but no flux-change. \\
(4) Adding residual shear bias: on top of data set (3), we further include reduced shear, shear multiplicative bias and additive bias \cite{heymans2006,zuntz2018,li2021noise}, so that $e=(1+m)\gamma/(1-\kappa)+e^I+\mathcal{N}(0,\sigma_e^2)+c$. The amplitudes of the shear biases $m$ and $c$ are redshift-dependent, with larger biases at higher-$z$ to mimic the lower imaging quality.\\
(5) Adding residual mean-$z$ bias: on top of data set (4), a biased $z_{\rm est}$ that represents the residual after redshift calibration \cite{wright2025z} is used instead of the true-$z$ in $n(z)$ for the theoretical calculation. Similar to the shear bias, $z_{\rm est}$ is more biased at higher-$z$.\\
(6) Removing magnification bias: on top of data set (5), remove the cosmic magnification while keeping all the other biases.

Considering that the image of a galaxy at higher-$z$ is generally smaller and dimmer, with lower S/N, the measured shear and calibrated redshift normally contain larger biases. We adopt $m(z)=10^{-3}(1+z)^2$, $c_1(z)=c_2(z)=10^{-4}(1+z)$, and $z_{\rm est}=z+10^{-3}(1+z)^2$, which are visualized in appendix~\ref{app:cate}. Such choices are based on the expected level of residual systematics for Stage-IV surveys \cite{yao2023csst}, so that on average the biases are within the requirement, but with larger biases at higher-$z$ to test the algorithm stability.

% ===========================================================================
\section{Lensing ratio: theory, two-point and field-level estimator}
\label{sec:methods}
% ===========================================================================

\subsection{Theory}
\label{sec:ratio}

For a lens at redshift $z_l$ and a source at $z_s>z_l$, the amplitude of the
lensing signal is proportional to the geometric factor
\begin{equation}
\beta(z_l,z_s)=\frac{D_A(z_l)\,D_A(z_l,z_s)}{D_A(z_s)}\,,
\label{eq:beta}
\end{equation}
where $D_A$ is the angular-diameter distance and $D_A(z_l,z_s)$ is the
angular-diameter distance between the lens and the source.  For a lens bin
$L$ and a source bin $S$ with normalized redshift distributions
$n_L(z_l)$ and $n_S(z_s)$, the appropriate average is
\begin{equation}
\ap{\beta}_{LS}
=
\frac{\displaystyle
\int\!\dd z_l\, n_L(z_l)
\int\!\dd z_s\, n_S(z_s)\,
\beta(z_l,z_s)\,\Theta(z_s-z_l)}
{\displaystyle
\int\!\dd z_l\, n_L(z_l)
\int\!\dd z_s\, n_S(z_s)}\,,
\label{eq:betaavg}
\end{equation}
with $\Theta$ the Heaviside step function.  The geometric prediction for the
ratio is
\begin{equation}
\Rgeo(L;S_1,S_2)
=
\frac{\ap{\beta}_{L,S_2}}{\ap{\beta}_{L,S_1}}\,.
\label{eq:Rgeo}
\end{equation}
When systematics vanish, this prediction depends only on the cosmology and on the measured $n(z)$ of
the three bins. For narrow redshift bins, this ratio can reduce to the mean-redshift approximation
\begin{equation}
\Rmez(L;S_1,S_2)
=
\frac{\beta(\bar z_l,\bar z_{s,2})}{\beta(\bar z_l,\bar z_{s,1})}\,.
\label{eq:Rmez}
\end{equation}

\subsection{The tangential-shear two-point ratio}
\label{sec:2pcf}

In two-point statistics, the lensing ratio is defined from the angular cross-power spectrum between the lens galaxy density and the source convergence. In the Limber approximation,
\begin{equation}
	C_\ell^{g\kappa}
	=
	\int \frac{\dd\chi}{\chi^2}\,
	W_g(\chi)\,W_\kappa(\chi)\,
	b_g(k,z)\,P_\delta(k,z)\Big|_{k=\ell/\chi}\,,
	\label{eq:Climber}
\end{equation}
where $W_g(\chi)=n_L(z)\,\dd z/\dd\chi$ is the normalized lens redshift distribution, $b_g(k,z)$ is the non-linear galaxy bias, $P_\delta(k,z)$ is the non-linear matter power spectrum, and $W_\kappa(\chi)$ is the lensing efficiency of the source bin. For a normalized source distribution $n_S(z)$,
\begin{equation}
	W_\kappa(\chi)
	=
	\frac{3H_0^2\Omega_m}{2c^2}\,
	\frac{\chi}{a(\chi)}
	\int_{z(\chi)}^{z_h}\dd z'\,
	n_S(z')\,
	\frac{\chi(z')-\chi}{\chi(z')}\,.
	\label{eq:Wkappa}
\end{equation}
Here $z_h$ is the horizon redshift, and the lensing ratio for two source bins behind the same lenses is then
\begin{equation}
	R_{2\rm pt}(L;S_1,S_2|\ell)
	=
	\frac{C_\ell^{g\kappa_{S_2}}}{C_\ell^{g\kappa_{S_1}}}
	=
	\frac{\int \dd\chi\,\chi^{-2}\,
		W_g(\chi)\,W_{\kappa,S_2}(\chi)\,
		b_g(k,z)\,P_\delta(k,z)}
	{\int \dd\chi\,\chi^{-2}\,
		W_g(\chi)\,W_{\kappa,S_1}(\chi)\,
		b_g(k,z)\,P_\delta(k,z)}\,.
	\label{eq:R2pt}
\end{equation}
Since $W_g$, $b_g(k,z)$, and $P_\delta(k,z)$ are common to the two source
bins, they cancel in the ratio for sufficiently narrow bins, so the ratio is
controlled by the lensing efficiency. For a single source at $z_s$,
$W_\kappa(\chi_l)\propto\beta(z_l,z_s)$, so for narrow source bins
\begin{equation}
	R_{2\rm pt}(L;S_1,S_2)
	\simeq
	\frac{\ap{\beta}_{L,S_2}}{\ap{\beta}_{L,S_1}}
	=
	\Rgeo(L;S_1,S_2)\,.
	\label{eq:R2pt_geo}
\end{equation}
In reality, the widths of the photo-$z$ bins are generally non-negligible. Therefore, Eq.\,\eqref{eq:R2pt} will end up being weakly scale- (i.e., $\ell$- or $\theta$-) dependent \cite{sun2023}, making the lensing ratio not a pure geometric probe, but also weakly dependent on the large-scale structure. Later we will show this scale-dependence is negligible when the bins are separated far enough, so that the measurement of the ratio can be compressed into a single ratio, given appropriate covariance.

We further note Eq.\,\eqref{eq:R2pt_geo} does not hold exactly, as in Eq.\,\eqref{eq:R2pt} the non-linear galaxy bias $b_g(k,z)$ and power spectrum $P_\delta(k,z)$ act as additional weights, which overall reshape the impact from $\chi$ \cite{sun2023}.

We measure $\gamma_t$ with
\textsc{TreeCorr} \cite{jarvis2004} in ten logarithmic bins from $\theta=0.1$ to $100\arcmin$. The ratio in each angular bin is
$R(\theta_i)=\gamma_{t,2}(\theta_i)/\gamma_{t,1}(\theta_i)$.  Its covariance
is obtained from the same 200-patch jackknife partition used for all other
estimators: for each patch we form the leave-one-patch-out shears
$\gamma_{t,1}^{(j)}$ and $\gamma_{t,2}^{(j)}$ and the corresponding ratios
$R_j(\theta_i)$.  The scalar ratio is the generalised least-squares (GLS)
constant
\begin{equation}
R_{2\rm pt}
=
\frac{\mathbf 1^{\rm T}C^{-1}\mathbf R}
     {\mathbf 1^{\rm T}C^{-1}\mathbf 1}\,,
\qquad
\sigma_{2\rm pt}^2
=
\frac{1}{\mathbf 1^{\rm T}C^{-1}\mathbf 1}\,,
\label{eq:gls}
\end{equation}
where $\mathbf R$ is the vector of $R(\theta_i)$ in the selected bins. This procedure is repeated for both the galaxy-shear correlation ratio and the galaxy-convergence correlation ratio.

\subsection{Convergence reconstruction with AKRA}
\label{sec:akra}

Our field-level estimators operate on convergence maps and thus require the reconstruction of convergence from the observed shear. Survey masks induce mode coupling and $E$/$B$-mode leakage, while spatially varying shape noise calls for proper inverse-variance weighting; a direct Kaiser--Squires inversion \cite{kaiser1993} does not account for either effect. We therefore adopt the Accurate Kappa Reconstruction Algorithm (AKRA) \cite{shi2024,shi2025,shi2026}, which explicitly incorporates the survey mask and noise covariance into the shear-to-convergence inversion without assuming any prior on the underlying field.\footnote{Other high-precision reconstructions typically incorporate log-normal priors \cite{fiedorowicz2022,boruah2024}, zero $B$-mode priors \cite{jeffrey2021}, or simulation-based priors \cite{jeffrey2020deepmass,white2026npe,han2026miu}, which would introduce external information that needs to be carefully considered when directly comparing $\gamma$ and $\kappa$.}

We describe the observed shear in harmonic space as
\begin{equation}
	\boldsymbol{y}=A\boldsymbol{x}+\boldsymbol{n},
\end{equation}
where $\boldsymbol{y}$ and $\boldsymbol{x}$ are the spherical-harmonic coefficients of the observed shear and the underlying convergence, respectively, and $\boldsymbol{n}$ is the noise vector with covariance $N$. The forward operator $A$ relates convergence to spin-2 shear and accounts for the mode coupling induced by the survey mask.

To stabilise the inversion, we add a quadratic regularisation term and solve the regularised least-squares problem
\begin{equation}
	\hat{\boldsymbol{x}}
	=
	\arg\min_{\boldsymbol{x}}
	\left[
	(A\boldsymbol{x}-\boldsymbol{y})^\dagger
	N^{-1}
	(A\boldsymbol{x}-\boldsymbol{y})
	+\lambda\,\boldsymbol{x}^\dagger\boldsymbol{x}
	\right],
	\label{eq:akra}
\end{equation}
where $\lambda$ controls the strength of the regularisation. The corresponding normal equations,
\begin{equation}
	\left(A^\dagger N^{-1}A+\lambda I\right)
	\hat{\boldsymbol{x}}
	=
	A^\dagger N^{-1}\boldsymbol{y},
\end{equation}
are solved using a matrix-free conjugate-gradient method \cite{shi2026}.

For the reconstruction, we build a shear map at $N_{\rm side}=4096$ by averaging galaxy ellipticities in each pixel and smoothing it with a Gaussian kernel of width $\sigma=1\arcmin$ (comparisons with other smoothing scales are presented in Appendix~\ref{app:smoothing}). After smoothing, the effective number of galaxies per pixel is given by $n_{\rm eff}=N_{\rm pix}\sigma^2 n_{\rm gal}$, and we mask pixels with $n_{\rm eff}<5$ due to their low S/N. In pixel space, we approximate the noise covariance as diagonal, with $N^{\rm pix}_{pp}=\sigma_e^2/n_{{\rm eff},p}$, and perform the reconstruction using $\ell_{\rm max}=8192$ and $\lambda=10^{-3}$.

\subsection{The field-level information}
\label{sec:fieldinfo}

A complete field-level lensing ratio will depend on a clear description of the likelihood $p(R|\vec{\kappa}_1,\vec{\kappa}_2,\vec{\delta}_g)$, which will require a complex suppression network and a large amount of realistic mocks to recover $\vec{\kappa}_{1/2}(\vec{\delta}_g)$ at each pixel. Here the vectors $\vec{\kappa}$ and $\vec{\delta}_g$ represent the full fields, while at pixel $i$ the values $\kappa_i$ and $\delta_{g,i}$ are correlated with the rest of the field. Simply assuming the pixels are uncorrelated will reduce the $\vec{\kappa}_{1/2}(\vec{\delta}_g)$ to the zero-lag two-point correlations \cite{wandelt2004gibbs,millea2020delensing,eriksen2004ilc,cardoso2008smica}. However, since in Sec.\,\ref{sec:akra} a smoothing is used before converting $\gamma$ into $\kappa$, part of the non-Gaussian information of the field is converted to Gaussian, making the zero-lag correlation of the convolved field not purely Gaussian. This is similar to how the other Gaussianization methods (the marked correlation function, Copula, etc.) subtract the non-Gaussian information \cite{white2016marked,xiao2026marked,qin2020copula}.

Based on this, we introduce the following two estimators to subtract the incomplete field-level information from zero-lag correlations for all pixels, and we describe how to derive the lensing ratio information. Demonstrations of the extra information being subtracted are shown in Sec.\,\ref{sec:compare} and Sec.\,\ref{sec:lcdm}.

\subsection{The field-level slope ratio}
\label{sec:slope}

After converting the non-local $\gamma$ into the local $\kappa$, a field-level direct comparison can be obtained between the $\kappa=W_\kappa\delta_m$ field and the galaxy overdensity field $\delta_g=b_g\delta_m$, where $\delta_m$ is the matter overdensity at given $z$. We note that overdensities from other redshift ranges do not correlate with $\delta_m$ and thus contribute to $\kappa$ as noise. Given $\delta_g$ from a pixel, the two corresponding lensing signals follow $\kappa_2(\delta_g) = R\kappa_1(\delta_g)$ when systematics vanish. 

The slope ratio estimator fits
\begin{equation}
\kappa_s(\hat{\bf n}) = a_s + b_s\,n_{\rm lens}(\hat{\bf n}) + \epsilon_s(\hat{\bf n})\,,
\qquad s=1,2\,,
\label{eq:slopefit}
\end{equation}
on the common pixel set of the lens galaxy number density map $n_{\rm lens}(\hat{\bf n})$ \footnote{We note using overdensity $\delta_g=(n-\bar{n})/\bar{n}$ is effectively identical to using number density, as this is a linear transformation.} and the two convergence maps $\kappa_s(\hat{\bf n})$, by ordinary least squares \footnote{This is where the zero-lag approximation discussed in Sec.\,\ref{sec:fieldinfo} comes into use. As the ``single-pixel versus full-field'' correlation is unknown, ordinary least squares assumes each pixel is independent.}, and defines
\begin{equation}
\Rslope=\frac{b_2}{b_1}\,.
\label{eq:Rslope}
\end{equation}
The intercept $a_s$ absorbs the mean field (for both convergence and number density) and any constant additive contribution that is uncorrelated with $\delta_m$ (including galaxy stochastic bias, shear additive bias, etc.); the slope $b_s$ contains information including not only the lensing efficiency and hence the geometric factor (Eq.\,\eqref{eq:beta}), but also systematics correlated with $\delta_m$ (shear multiplicative bias, cosmic magnification, intrinsic alignment, etc.).

Considering non-linear galaxy bias, the slope of $b_s$ varies for different $\delta_g$, thus Eq.\,\eqref{eq:slopefit} holds only over a reasonably narrow range of $\delta_g$ and $\kappa$, while the linearity breaks for a wide range. However, because the ratio in Eq.\,\eqref{eq:Rslope} cancels the non-linear bias contribution in both the numerator and the denominator, the bias introduced by the linear galaxy bias assumption is expected to be reduced.

\subsection{The field-level direct ratio}
\label{sec:direct}

To avoid the linear galaxy bias assumption, we further apply the ratio estimator of Sun et al.\ \cite{sun2023}, which directly measures the ratio of two distributions $\kappa_2(\delta_g)/\kappa_1(\delta_g)$ while marginalizing over their possible common template $\kappa(\delta_g)$.  We
bin the common pixels in ten bins of $n_{\rm lens}$ between 0 and 20 and
form the $2n$-element data vector
\begin{equation}
{\bf d}=
\begin{pmatrix}{\bf d}_1\\{\bf d}_2\end{pmatrix}\,,
\qquad
{\bf d}_1={\boldsymbol\lambda}+{\bf n}_1\,,
\qquad
{\bf d}_2=R\,{\boldsymbol\lambda}+{\bf n}_2\,,
\label{eq:sunmodel}
\end{equation}
where ${\boldsymbol\lambda}$ is the unknown signal template, $R$ is the
ratio, and ${\bf n}_1,{\bf n}_2$ are Gaussian noise vectors with the joint
covariance
\begin{equation}
C=
\begin{pmatrix}
C_{11} & C_{12}\\
C_{21} & C_{22}
\end{pmatrix}\,.
\label{eq:suncov}
\end{equation}
The covariance is estimated from jackknife resampling, so it
includes the correlation between the two source bins.  Marginalising over
${\boldsymbol\lambda}$ analytically gives
\begin{equation}
-2\ln P(R\mid{\bf d})
=
\ln\det Q(R)
-
{\bf T}(R)^{\rm T}Q(R)^{-1}{\bf T}(R)
+
{\rm const}\,,
\label{eq:sunpost}
\end{equation}
with
\begin{equation}
Q(R)=B_{11}+R\ap{B_{12}+B_{21}}+R^2B_{22}\,,
\label{eq:sunQ}
\end{equation}
\begin{equation}
{\bf T}(R)
=
\ap{B_{11}{\bf d}_1+B_{12}{\bf d}_2}
+
R\ap{B_{21}{\bf d}_1+B_{22}{\bf d}_2}\,,
\label{eq:sunT}
\end{equation}
where $B_{ij}$ are the blocks of $C^{-1}$.  The maximum-posterior value and the 68\% credible interval are obtained by evaluating equation~\eqref{eq:sunpost} on a fine grid in $R$; the determinant term is retained. 

This estimator assumes only that the two data vectors are proportional: it makes no assumption about the shape of the signal and therefore no assumption about the galaxy bias. We consider this method as the proper way for the direct lensing ratio, while the slope ratio in Sec.\,\ref{sec:slope} acts as a comparison to show the potential bias from the linear galaxy bias assumption. Later in Sec.\,\ref{sec:fidfield} we discuss where the zero-lag approximation affects this estimator.

\subsection{Joint field+two-point estimator}
\label{sec:joint}

The field-level and two-point estimators use the same shear field but weight
the information differently.  We form the joint data vector
\begin{equation}
{\bf y}
=
\ap{\Rdirect,\,R(\theta_2),\,\ldots,\,R(\theta_9)}\,,
\label{eq:jointvec}
\end{equation}
estimate its covariance from the jackknife resampling, and combine the elements with the GLS weights of equation~\eqref{eq:gls}. The resulting estimator is denoted $\Rjoint$.  The joint uses the direct ratio rather than the slope ratio to avoid additional assumptions.

\subsection{Jackknife covariance and the Hartlap correction}
\label{sec:jk}

All estimators use the same 200-patch jackknife partition, obtained by
applying a K-means algorithm to the observed lens positions of the
lowest-redshift bin.  The partition is frozen after the first generation and
its checksum is verified at the start of every measurement.  For a data
vector ${\bf X}$ with $p$ elements and $n=200$ patches, the covariance is
\begin{equation}
C=\frac{n-1}{n}\sum_{j=1}^{n}
\ap{{\bf X}_j-\bar{\bf X}}\ap{{\bf X}_j-\bar{\bf X}}^{\rm T}\,.
\label{eq:jkcov}
\end{equation}
The inverse is biased because $C$ is estimated from a finite number of
patches.  We apply the Hartlap correction \cite{hartlap2007}
\begin{equation}
C^{-1}\;\longrightarrow\;
\frac{n-p-2}{n-1}\,C^{-1}\,.
\label{eq:hartlap}
\end{equation}
For the 35-triple fit ($p=35$, $n=200$) the factor is $0.819$.  The
jackknife covariance is itself noisy; with $p/n=0.175$ the inverse is noisier
still, and the Hartlap factor corrects the bias but not the variance
\cite{dodelson2013,sellentin2016,friedrich2018}.  This is particularly
relevant for the field-level estimator, whose covariance has a
near-degenerate mode that we discuss in section~\ref{sec:discussion}.

\subsection{Flat \texorpdfstring{$\Lambda$}{Lambda}CDM inference}
\label{sec:mcmc}

For each triple $t$ we measure a ratio $R_t$ and its covariance $C_{tt'}$
from the 35-triple jackknife.  The likelihood is
\begin{equation}
-2\ln\mathcal L(\Om)
=
\sum_{t,t'}
\bs{R_t-R_{\rm th,t}(\Om)}
\ap{C^{-1}}_{tt'}
\bs{R_{t'}-R_{\rm th,t'}(\Om)}\,,
\label{eq:like}
\end{equation}
with the theoretical prediction $R_{\rm th}$ given by equation~\eqref{eq:Rgeo} and the same $n(z)$
histograms used for the measurement.  We consider flat $\Lambda$CDM with
$\Om$ free and compute $R_{\rm th}$ from the flat-$\Lambda$CDM
angular-diameter distances.  $H_0$ cancels in the ratio and is fixed to
$67.66\ \rm km\,s^{-1}\,Mpc^{-1}$; $\Omega_b$ is fixed to $0.049$ and is not
sampled.  There is no nuisance parameter: every estimator is compared directly with the pure-geometry prediction. The impact of the residual systematics in Sec.\,\ref{sec:systematics} therefore appears directly as a bias in the final result.

We sample the posterior with \textsc{emcee} \cite{foremanmackey2013}, using
a flat prior $\Om\in[0.05,0.90]$. The chains are started from a coarse random search followed by a Nelder--Mead refinement of the $\chi^2$, and the walkers are initialised in a small Gaussian ball around the best fit. We check convergence with the integrated autocorrelation time, discard the burn-in, and thin the chains accordingly.

% ===========================================================================
\section{Measurements on the fiducial triple}
\label{sec:fiducial}
% ===========================================================================

We first apply the four estimators to the fiducial triple
($0.1$--$0.4$, $0.8$--$1.0$, $1.6$--$3.2$) in the
(systematics level 2) IA+noise sample, as described in Sec.\,\ref{sec:systematics}, because the non-Gaussian intrinsic shapes closely mimic real observations. No systematics other than IA are included.

\subsection{The tangential-shear two-point ratio}
\label{sec:fid2pt}

\begin{figure}[tbp]
\centering
\includegraphics[width=\textwidth]{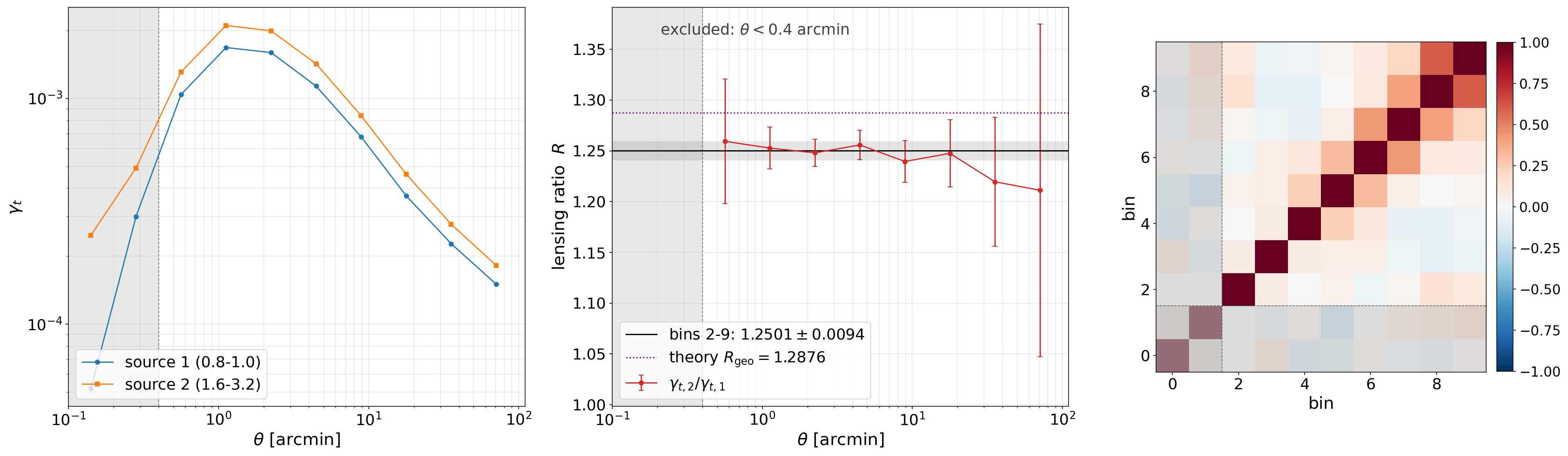}
\caption{Tangential-shear two-point measurements for the fiducial triple, and the resulting shear ratio. Left: tangential shear $\gamma_t(\theta)$ around the lenses in the $0.1$--$0.4$ bin for the two source bins. The shaded region and the dashed vertical line mark the two smallest angular bins, because the simulation is stored at $\rm nside=8192$ in \healpix. We therefore exclude $\theta<0.4\arcmin$ in all the following results. Middle: ratio $R(\theta)=\gamma_{t,2}/\gamma_{t,1}$. The red measurements do not deviate from a scale-independent ratio, which is shown in the solid horizontal line $R_{2\rm pt}=1.2501\pm0.0094$; the dotted line is the
$n(z)$-weighted geometric prediction $\Rgeo=1.2876$. Right: correlation matrix of the ratio data vector, with the excluded bins shaded.}
\label{fig:2pt}
\end{figure}

Figure~\ref{fig:2pt} shows the tangential shear for the two source bins, the ratio $R(\theta)$, and the correlation matrix of the ratio. The signal is detected over three orders of magnitude, from $0.1\arcmin$ to $100\arcmin$, divided into 10 logarithmic bins. We remove the two bins with $\theta<0.4\arcmin$ due to the limitation of the simulation, and find that the shear ratio is generally constant over bins
2--9 (removing bins 0 and 1). The correlation matrix in the right panel of figure~\ref{fig:2pt} shows the
expected block structure: adjacent angular bins are positively correlated, and the correlation increases with separation. Below we refer to this estimator as the 2PCF ratio. 

From the tangential-shear measurements and the covariance, a scale-independent ratio $R=1.2501\pm0.0094$ can be achieved. Due to the good linearity and high S/N of the two-point measurements, this ratio remains well behaved and Gaussian, and offers stronger constraining power than any individual angular bin. However, the constant ratio is $2.9\%$ below the $n(z)$-weighted prediction $\Rgeo=1.2876$ but within $0.36\%$ of the mean-redshift approximation $\Rmez=1.2546$. This emphasizes that when the S/N of the ratio is high enough, theoretical modeling becomes important. We return to this point in Sec~\ref{sec:discussion}.

\subsection{The field-level ratios: slope \& direct}
\label{sec:fidfield}

\begin{figure}[tbp]
\centering
\includegraphics[width=\textwidth]{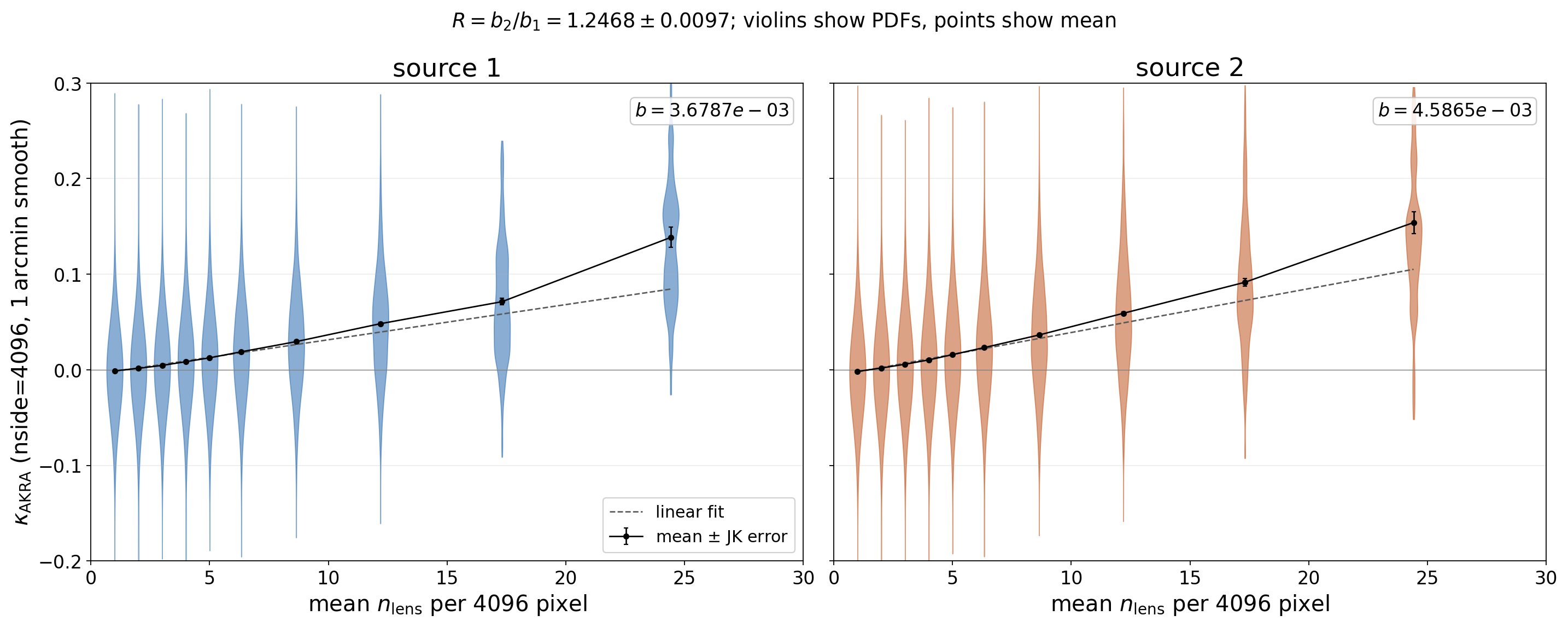}
\caption{Field-level convergence--(galaxy number density) relation for the fiducial triple.  Left: source bin $0.8$--$1.0$; right: source bin $1.6$--$3.2$. The maps are reconstructed by AKRA at $\rm nside=4096$ with $1\arcmin$ Gaussian smoothing on the shear maps. We divide the galaxy number density into 10 bins and estimate the mean $\kappa$ in each bin, with jackknife error bars. The violins show the per-pixel
distribution, as the scatter diagram itself is quite noisy. The dashed line is the linear fit with Eq.\,\eqref{eq:slopefit}, which agrees with the black mean values at $\kappa<\sim0.3$ and $n_{\rm lens}\le7$. Although the linear galaxy bias assumption breaks at high-density regions, the linear fit is still dominated by the regions with lower density as there are more galaxies. The fitted slope ratio is $\Rslope=b_2/b_1=1.2468\pm0.0097$. The direct ratio of Eq.\,\eqref{eq:sunpost} calculated from the black dots gives $\Rdirect=1.2456\pm0.0097$, which is consistent with $\Rslope$.}
\label{fig:field}
\end{figure}

Figure~\ref{fig:field} shows the relation between the AKRA-reconstructed convergence $\kappa$ maps (with $1\arcmin$ smoothing on the $\gamma$ maps) and the lens galaxy number counts at $\rm nside=4096$. Points are the mean convergence in ten bins of the mean lens number count per pixel, with jackknife error bars; the violins show the per-pixel distribution; the dashed line is the linear fit with an intercept. The relation is well described by a straight line over
$n_{\rm lens}\lesssim7$, while at $n_{\rm lens}>7$ there is an apparent effect due to non-linear galaxy bias. The fitted slope ratio is $\Rslope=1.2468\pm0.0097$, with an error of about $0.8\%$, comparable to the two-point error. The central value differs from the two-point scalar by $0.0033$, well within the errors.

The scatter of the violins in Fig.\,\ref{fig:field} around the mean values is large: this is not only due to shape noise, but also because $\kappa$ is correlated with $\delta_g$ or $n_{\rm lens}$ only at the same $z$, while at a different $z$ the matter field $\delta_m$ is uncorrelated with the lens $\delta_g$, making the corresponding $\kappa$ contribute as random noise. Nonetheless, as many pixels carry the $\kappa-\delta_g$ correlation, their mean values together suppress the noise and form a semi-linear relation with very high S/N. More discussion on the cosmic information will be shown in the following subsection.

Meanwhile, when directly calculating the ratio of the mean $\kappa$ values in Fig.\,\ref{fig:field}, the result gives $\Rdirect=1.2456\pm0.0097$, with a central value close to $\Rslope$ and comparable scatter. This seems to suggest the linear galaxy bias assumption does not bias the ratio as the numerator and denominator cancel. We will show later that this argument does not always hold. Still, as most galaxies reside in less dense regions, the linear fitting agrees very well at $n_{\rm lens}\le7$, while at the massive end, the violin PDFs are multi-peak, due to limited pixel samples (it might also be related to assembly bias, but that is beyond the scope of this paper). 

We note that, when calculating the mean $\kappa$ values in each bin in Fig.\,\ref{fig:field}, each pixel on the map is assumed to be independent, so that the zero-lag approximation (Sec.\,\ref{sec:fieldinfo}) applies. Many pixels in a bin build up the violin distribution. Therefore, this method also presents part but not full information from the field.

\subsection{Shared information between the estimators}
\label{sec:shared}

\begin{figure}[tbp]
\centering
\includegraphics[width=\textwidth]{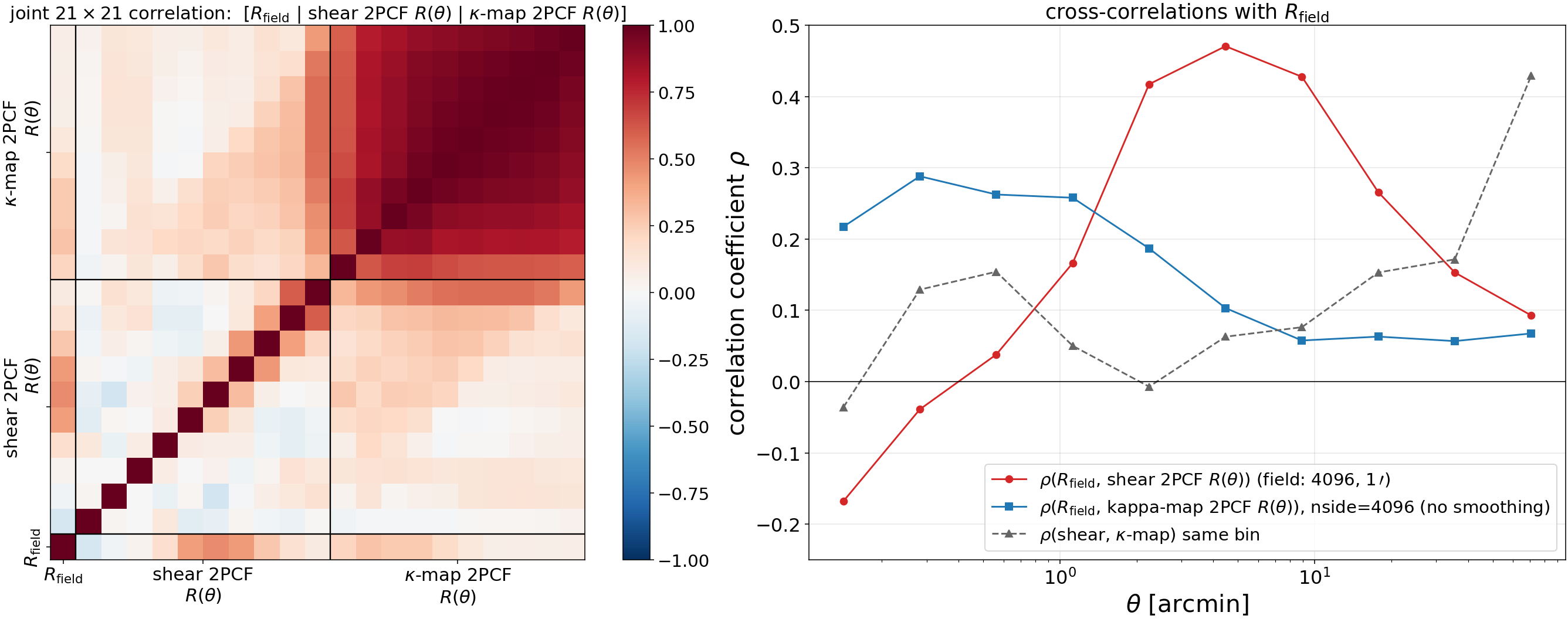}
\caption{Shared information (normalized covariance matrix) between the direct ratio and the two-point estimator. Left: $21\times21$ correlation matrix of $[\Rdirect$, shear 2PCF $R(\theta)$ (10 bins), $\kappa$-map 2PCF $R(\theta)$ (10 bins)$]$, estimated from the 200-patch jackknife. Right: cross-correlation coefficients with $\Rdirect$ as a function of $\theta$, which correspond to the bottom line of the left panel. The shared information between the direct ratio and the shear 2PCF ratio peaks at $\sim5$ arcmin (red), while it peaks at <1 arcmin for the direct ratio and the $\kappa$ 2PCF (blue). This demonstrates that the reconstructed $\kappa$ field is local, and its direct comparison with $\delta_g$ in the same pixel can capture a majority of the field information, given that nearby $\kappa$ values are strongly correlated. We use $R_{\rm field}$ to denote the direct ratio $\Rdirect$ as the field-level entry; replacing it with the slope ratio $\Rslope$ gives a visually identical covariance, however, they differ in detail (Appendix \ref{app:cov}).
}
\label{fig:shared}
\end{figure}

To combine the field-level and two-point ratios as described in Sec.\,\ref{sec:joint}, a net covariance is needed. This also helps us understand where the information comes from and perform sanity checks. We construct a $21$-element data vector from $\Rdirect$, the ten shear ratios, and the ten $\kappa$-map ratios, and estimate its $21\times21$ covariance from the 200 patches. The normalized covariance (i.e., correlation coefficient) is shown in Fig.\,\ref{fig:shared}. 

In the right panel of Fig.\,\ref{fig:shared}, we see that the direct ratio is moderately correlated with the shear two-point ratio over a wide angular range (red), with $\rho\simeq0.04$--$0.48$ and a maximum at $\theta\simeq4$--$5\arcmin$. This demonstrates that AKRA successfully concentrates the widely-spread non-local information from the $\gamma$ field into the local $\kappa$ field, allowing our algorithm to estimate the field-level ratio, and the peak exhibits the balance between correlation strength and noise. Meanwhile, its correlation with the $\kappa$-map two-point ratio peaks at the center (blue), suggesting that the main information of the $\kappa$ field resides within the pixel. The shear and $\kappa$-map ratios are almost uncorrelated at small scales with the same $\theta$ and become moderately correlated at large
$\theta$. This further illustrates that AKRA reconstructs the local $\kappa$ from the non-local $\gamma$. This result suggests that the direct ratio shares part of its information
with the 2PCF ratios. But the correlation coefficient is far from unity, making it possible to extract more information from the combination of the two.

We therefore combine the $\Rdirect$ and $R_{2\rm pt}$ from the shear 2PCF, and an enhanced $\Rjoint=1.2486\pm0.0087$ is achieved. This is about $7\%$ better than the $R_{2\rm pt}$ case.

\subsection{Comparison of different estimators}
\label{sec:compare}

\begin{figure}[tbp]
\centering
\includegraphics[width=0.86\textwidth]{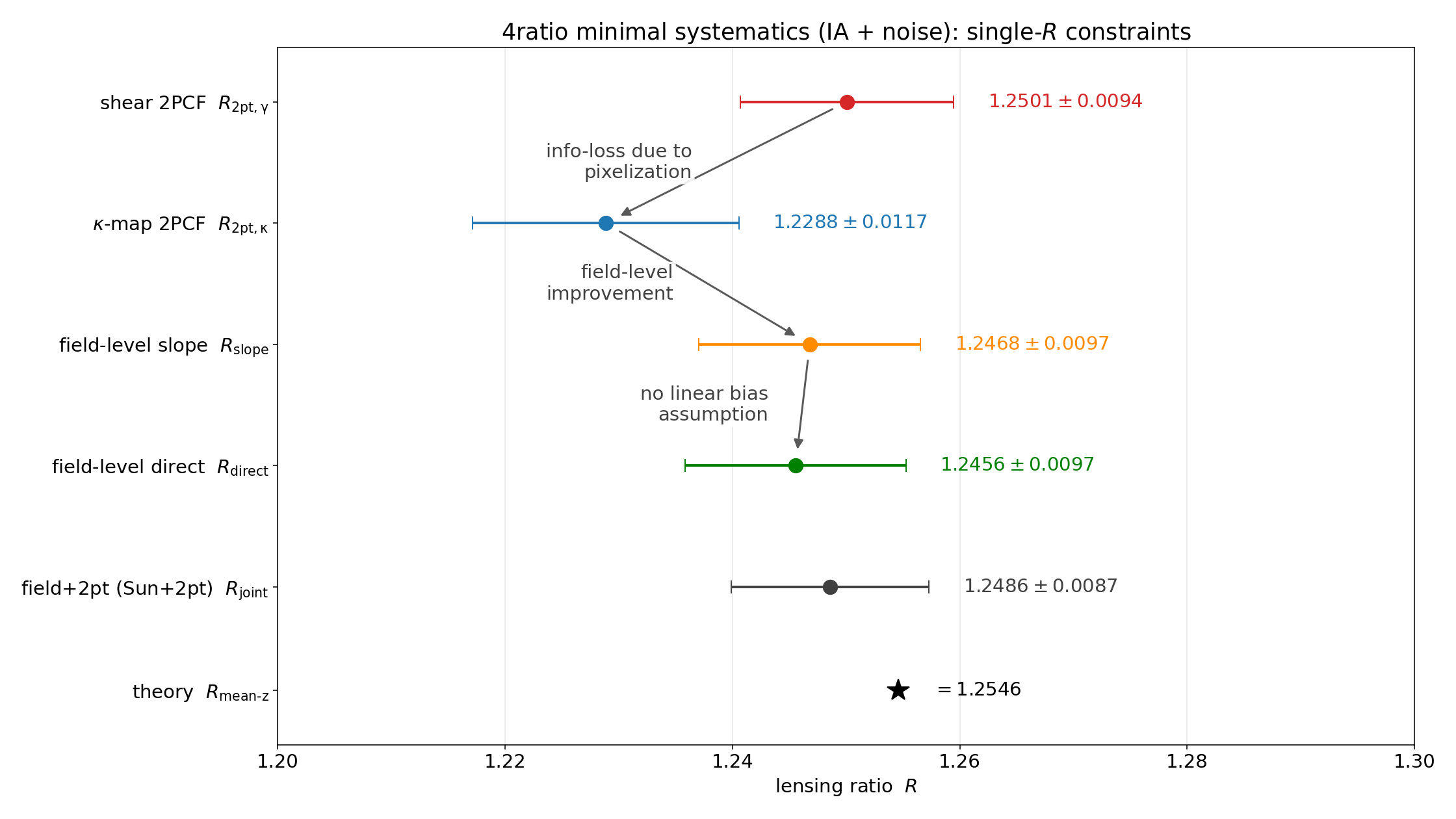}
\caption{Comparisons of different ratio estimators for the fiducial triple.  From top to bottom: shear two-point ratio $R_{2\rm pt,\gamma}$
(bins 2--9, with bins 0--1 removed, same as Fig.\,\ref{fig:2pt}), unsmoothed $\kappa$-map two-point ratio $R_{2\rm pt,\kappa}$, the field-level slope ratio $\Rslope$ (Fig.\,\ref{fig:field}), the field-level direct ratio $\Rdirect$, and the joint field+two-point ratio $\Rjoint$. The arrows indicate the information flow during the transfer. The star is the theoretical prediction using mean-redshift geometry $\Rmez=1.2546$. The impact of different theoretical predictions will be discussed later.}
\label{fig:compare}
\end{figure}

Figure~\ref{fig:compare} collects the single-value ratios from the different estimators for the fiducial triple and compares how the information is transferred:\\
$\bullet$ The conventional shear 2PCF ratio is $R_{2\rm pt,\gamma}=1.2501\pm0.0094$, calculated directly from the shear catalog.\\
$\bullet$ After converting the galaxy shear catalog into $\rm nside=4096$ ($\sim0.74$ arcmin$^2$/pixel) shear map, the sub-grid information is lost. As discussed before, AKRA offers the SOTA convergence-reconstruction without assuming any prior on the field, thus this step is optimal for the current stage. The $\kappa$ 2PCF ratio using this convergence map exhibits the information loss during the whole process, with $R_{2\rm pt,\kappa}=1.2288\pm0.0117$.\\
$\bullet$ By applying the field-level slope ratio estimator in Sec.\,\ref{sec:slope}, which effectively assumes linear galaxy bias, more field-level information is extracted, giving $\Rslope=1.2468\pm0.0097$. This is a significant improvement over $R_{2\rm pt,\kappa}$, demonstrating that this estimator successfully obtains non-Gaussian information, which helps reduce the errors in the numerator and denominator of the ratio compared with the 2PCF.\\
$\bullet$ By further applying the field-level direct ratio estimator in Sec.\,\ref{sec:direct}, no assumptions are made on the $\kappa(\delta_g)$ relation, so that an unbiased estimation of the ratio is obtained, with $\Rdirect=1.2456\pm0.0097$. Although this estimate is almost identical to $\Rslope$, it will be more biased in other cases, while the covariance of $\Rslope$ is more biased.\\
$\bullet$ Finally, by combining the $R_{2\rm pt,\gamma}$ and $\Rdirect$, together with the covariance part in Fig.\,\ref{fig:shared}, an improved value is obtained $\Rjoint=1.2486\pm0.0087$.

On the observational side, we see that the field-level ratio can extract more information from a convergence map with finite pixel size. However, sub-pixel information can only be obtained at the catalog level, making the shear 2PCF unique. The joint estimator $\Rjoint$ seems promising in this data set with minimal systematics (only IA), but the stability of the different estimators over various types of systematics still needs to be checked. 

On the theoretical side, when the three bins are widely separated, $\Rmez$ seems to offer a good description of the data. But when the binning changes, the agreement between measurements and the theoretical estimation also needs further examination. This leads to the next section, where we compare different estimations and predictions over different data sets and all 35 triples from the seven tomographic bins.

\raggedbottom

% ===========================================================================
\section{The redshift-bin scan and the systematics ladder}
\label{sec:scan}
% ===========================================================================
In this section, we gradually add systematics step-by-step, and check their impact on different ratio estimators at different triples. As discussed in Sec.\,\ref{sec:systematics}, the sequence of systematics added is: no systematics (pure Gaussian noise) => add IA (and the associated non-Gaussian intrinsic shape) => add lens magnification => add residual shear bias => add residual mean-$z$ bias.

\begin{figure}[tbp]
	\centering
	\includegraphics[width=\textwidth]{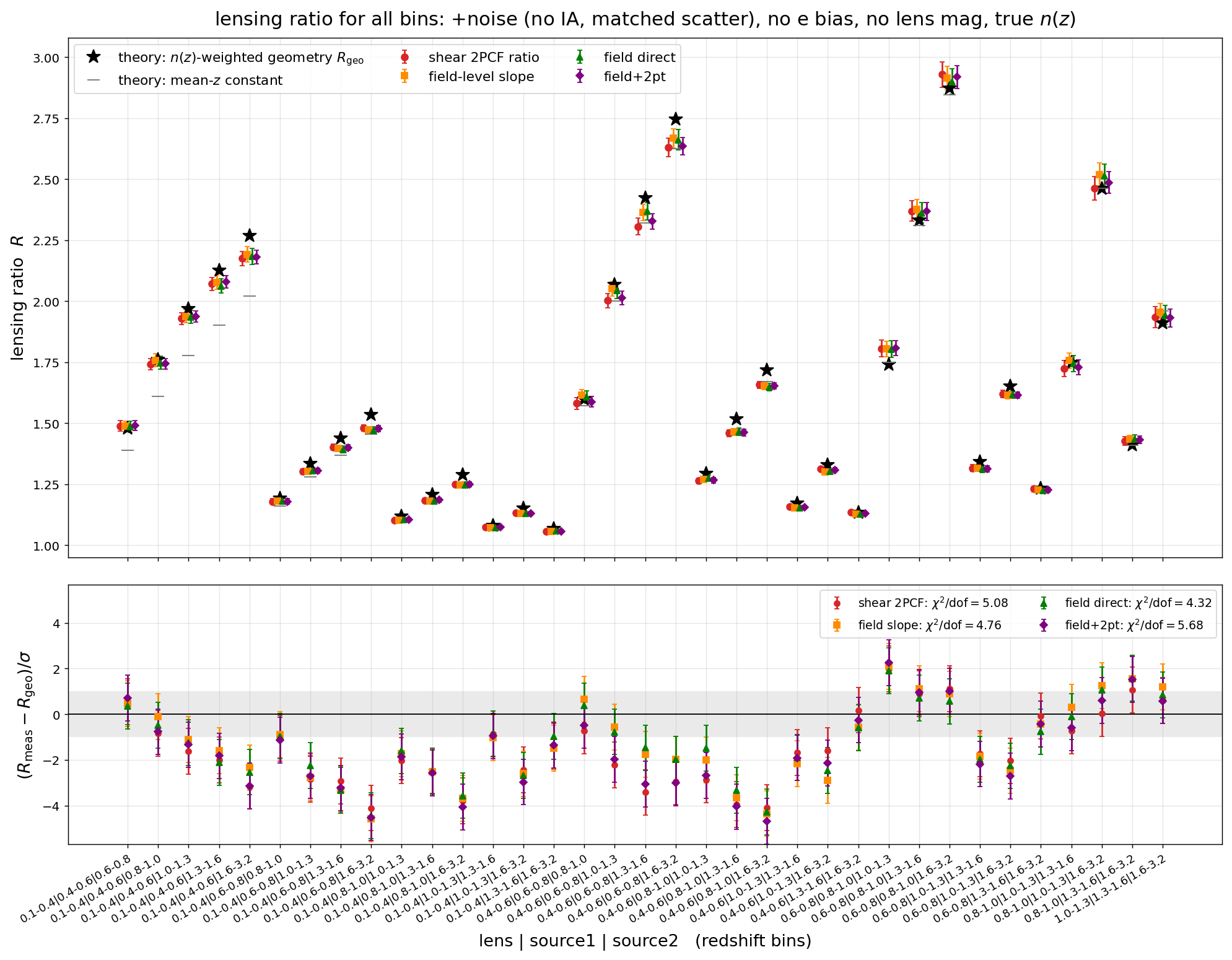}
	\caption{Lensing ratio for all 35 triples in the cleanest sample (level 1: Gaussian shape noise only, no intrinsic alignment, unlensed lens positions, no shear bias, true-$z$ generated $n(z)$). Top: measurements from four ratio estimators and two theoretical predictions. Black stars are $\Rgeo$; grey dashes are $\Rmez$; red measurement is $R_{2\rm pt}$; orange measurement is $\Rslope$; green measurement is $\Rdirect$; purple measurement is $\Rjoint$. Bottom: residual bias of $(R_{\rm measurement}-\Rgeo)/\sigma$; the grey band is $\pm1\sigma$. When observational systematics vanish, the measurements from all four estimators agree with each other, and their small error bars make the theoretical prediction the dominant source of bias. The measurements track the geometric trend from $R\simeq1.06$ to $3.0$ but lie systematically below $\Rgeo$ by about $1.5\%$, which leads to large reduced-$\chi^2$ values.}
	\label{fig:noia}
\end{figure}

\subsection{Gaussian shape noise only}
\label{sec:noia}

In the case of no observational systematics, we test the self-consistency between different estimators and their agreement with the theoretical predictions, which is similar to Fig.\,\ref{fig:compare} but over all the triples. Fig.\,\ref{fig:noia} shows the cleanest sample (level 1): Gaussian shape noise with the intrinsic alignment removed, unlensed lens positions, no residual shear systematics and the true-$z$ $n(z)$. 

We find that the four measurements are generally self-consistent; however, the two theoretical predictions deviate from the measurements at different circumstances: $\Rmez$ is significantly biased when the lens bin is 0.1--0.4 and the source1 bin is 0.4--0.6, while $\Rgeo$ is biased when the lens bin and source1 bin are adjacent. We see that the good agreement between data and model in Fig.\,\ref{fig:compare} only happens when the bins are widely separated. The reduced-$\chi^2$ between the theory and the data is large, with $\Rdirect$ at $4.32$ being the smallest and $\Rjoint$ at 5.68 being the largest. This indicates that the main limitation for a Stage-IV survey is the accuracy of modeling the ratio.

Because of the large residuals in Fig.\,\ref{fig:noia}, we note the current inaccuracy in the ratio modeling could limit the final output in cosmology. It could potentially be resolved by using many realistic mocks and building a field-level lensing ratio emulator, similar to what has been done in two-point statistics~\cite{chen2025emulator1,chen2025emulator2,zhou2025emulator3}.

\subsection{Adding intrinsic alignments}
\label{sec:ia}

\begin{figure}[tbp]
\centering
\includegraphics[width=\textwidth]{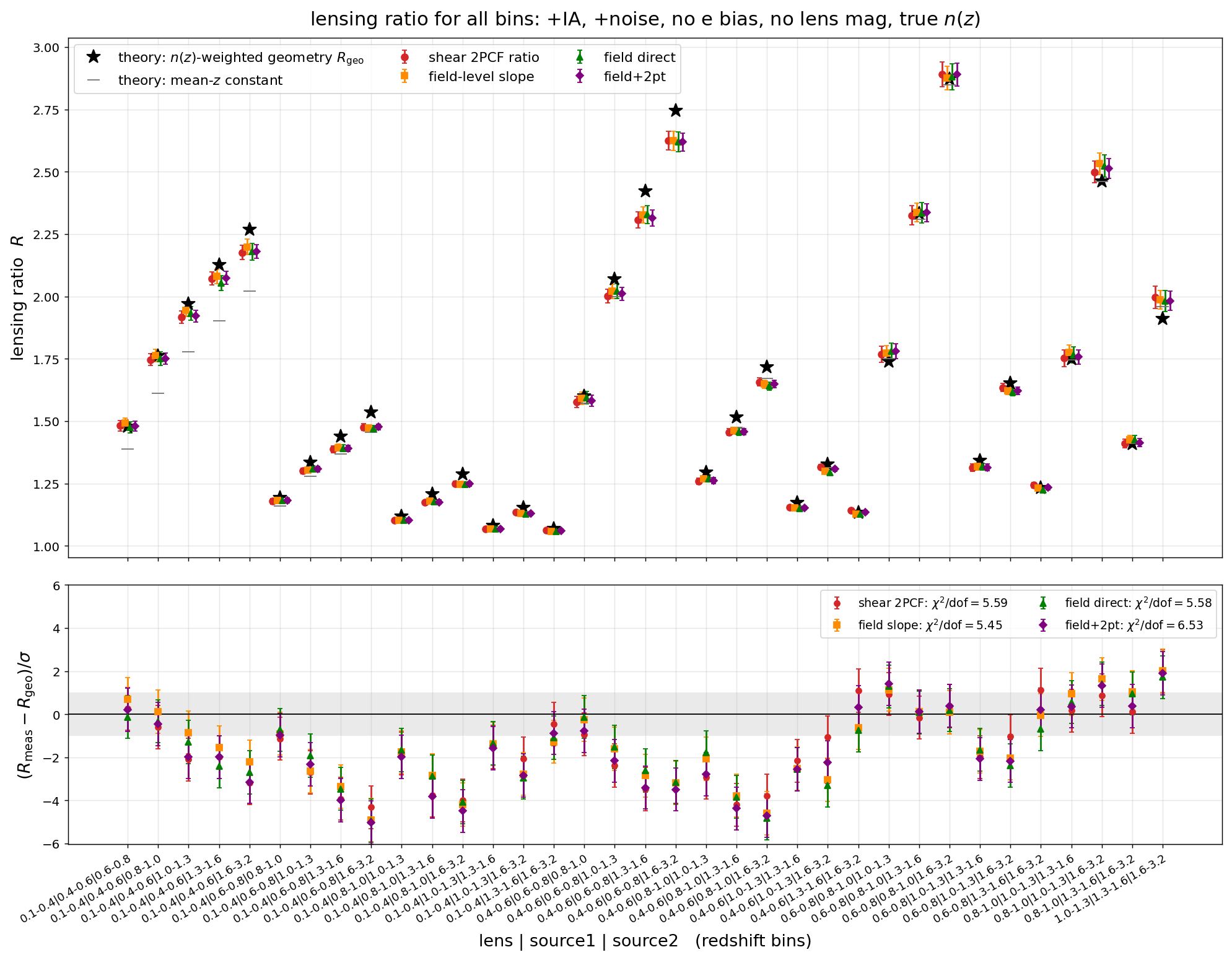}
\caption{Same as figure~\ref{fig:noia}, but for level 2: adding IA with non-Gaussian intrinsic shape, and Gaussian noise to control the total ellipticity rms at $0.275$. The addition of intrinsic alignments changes the ratios by less than about
$0.3\%$, demonstrating IA does not introduce a severe bias in the lensing ratio studies.}
\label{fig:ia}
\end{figure}

Fig.\,\ref{fig:ia} shows level 2: intrinsic alignments added, together with the galaxy non-Gaussian intrinsic shapes (an additional Gaussian shape noise is added so that the total ellipticity rms is unchanged). The measurements are almost indistinguishable from Fig.\,\ref{fig:noia}.  The change relative to level 1 is below about $0.3\%$ for all four estimators, smaller than the statistical error of a
single triple. We therefore conclude intrinsic alignments (at least the IA in our simulation) are not a severe source of systematics. This is most likely because the separated photo-$z$ bins already reject a majority of the galaxy-IA correlation in galaxy-galaxy lensing, and the field-level ratio is affected in a similar way. The reduced-$\chi^2$ increases slightly compared with Fig.\,\ref{fig:noia}, most likely due to the change in the noise properties.

\subsection{Adding lens magnification}
\label{sec:lensmag}

\begin{figure}[tbp]
	\centering
	\includegraphics[width=\textwidth]{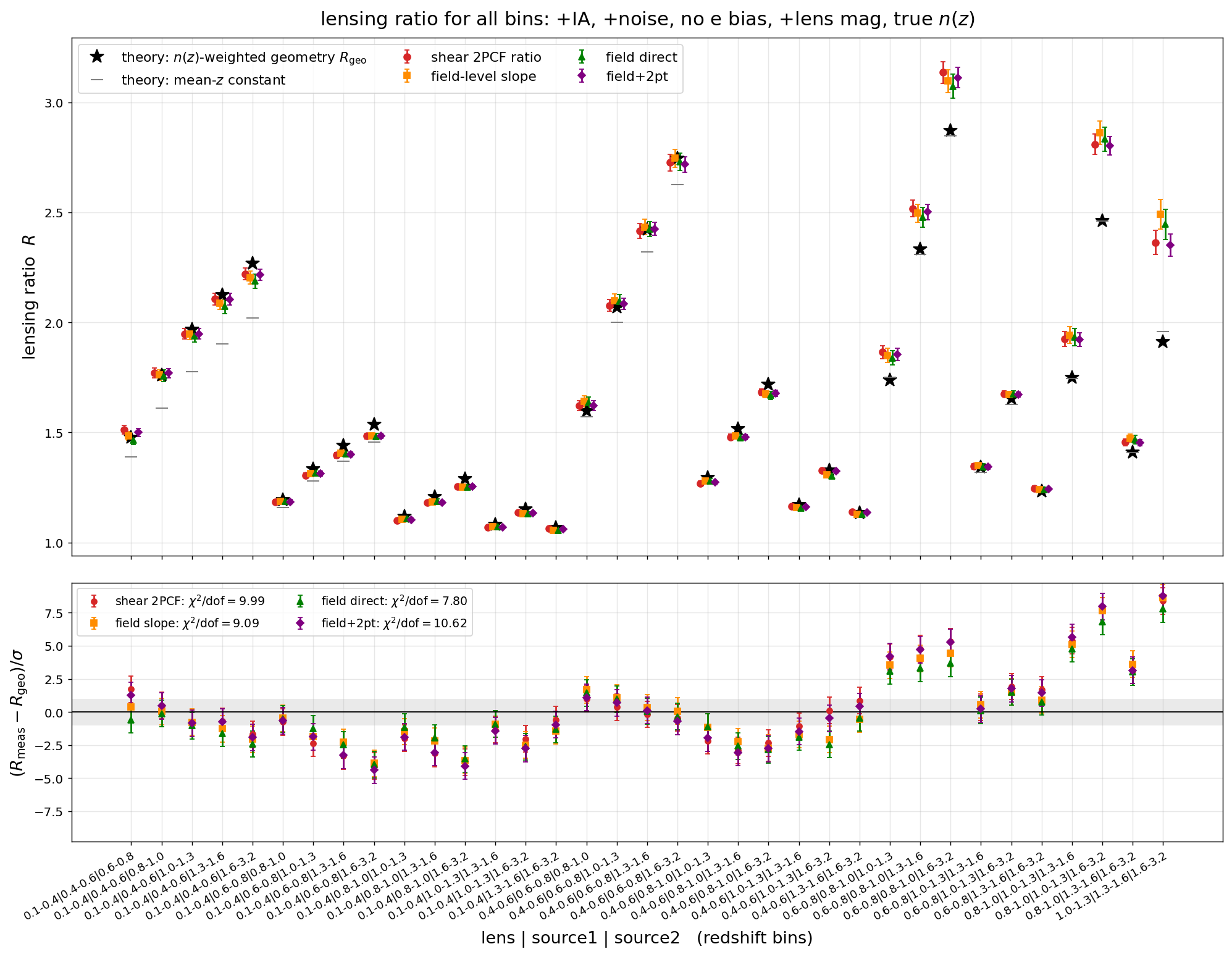}
	\caption{Same as figure~\ref{fig:noia}, but for level 3: besides IA and shape noise, the lens magnification is added. For the three high-$z$ lens bins, starting from the bin with photo-z in 0.6--0.8, a significant increase is found in the measurements for all ratio estimators. This leads to a large bias from the theoretical prediction and an inflated reduced-$\chi^2$.}
	\label{fig:mag}
\end{figure}

Fig.\,\ref{fig:mag} shows level 3: an additional bias from cosmic magnification of the lens is added. At low-$z$, lens magnification has a weak contribution to the measured ratio, but at high-$z$, the contamination from lensing magnification can strongly bias the ratio, as the numerator and denominator contain the contribution from magnification very differently. We see that in bottom panel of Fig.\,\ref{fig:mag} the residual bias, the left part remains similar to the previous Fig.\,\ref{fig:noia} and \ref{fig:ia}, while the right part (where the high-$z$ lenses reside) significantly increases. Consequently, the reduced-$\chi^2$ is significantly enlarged. This agrees with previous findings that cosmic magnification is non-negligible at high-$z$ \cite{yao2023kids1000}. In this work, this effect is more severe, possibly due to smaller statistical error and different sensitivity in the numerator and denominator; a direct illustration for one triple is shown in appendix~\ref{app:decomp}.

We note in our mock data, we only include the magnification from the area-change, so that the prefactor $g_{\rm mag}=-2$ and the effect from flux-change is not included. In real observations the $g_{\rm mag}$ could also change for different redshift bins. It is possible to suppress or even remove lens magnification with proper weighting \cite{qin2024,qin2025}, but its field-level performance is untested.

\subsection{The full systematics ladder}
\label{sec:alleff}

\begin{figure}[tbp]
\centering
\includegraphics[width=\textwidth]{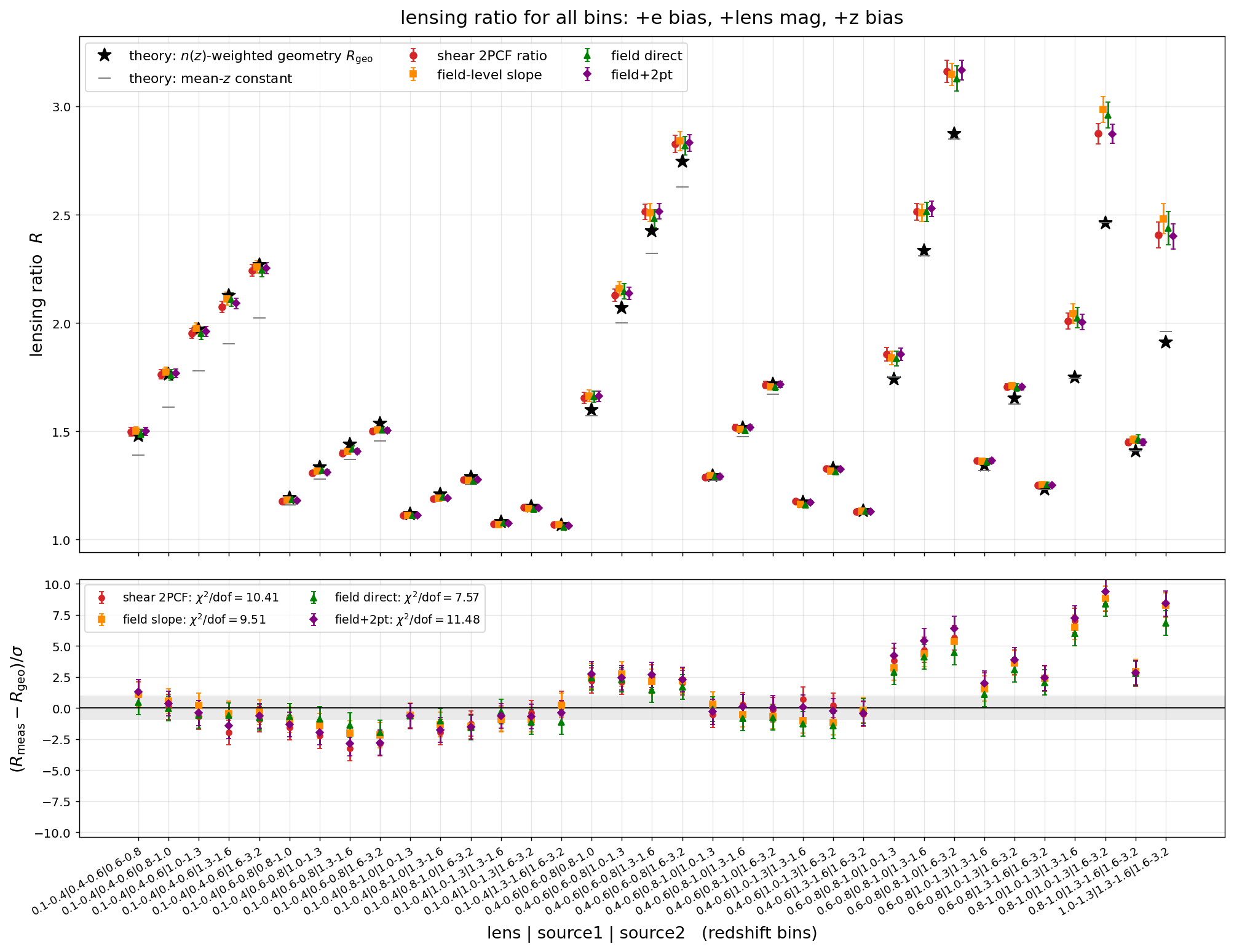}
\caption{Same as figure~\ref{fig:noia}, but for level 5: except for the IA and lens magnification being added in Fig.\,\ref{fig:mag}, residual shear bias and mean-$z$ bias (appendix~\ref{app:cate}) are also included. The ratios from the four different estimators change slightly from Fig.\,\ref{fig:mag}, as well as their reduced-$\chi^2$. This suggests shear multiplicative bias, reduced shear, and mean-$z$ bias overall contribute less impact than the lens magnification or the modeling error.}
\label{fig:alleff}
\end{figure}

We finally show the net contribution from residual shear bias and mean-$z$ bias (level 5), skipping the results of adding only the residual shear bias (level 4), because its impact is relatively small, similar to that of the mean-$z$ bias. The measurements in Fig.\,\ref{fig:alleff} show similar deviations from the theoretical predictions as the previous Fig.\,\ref{fig:mag}. Here we assume that the residual biases in shear and redshift are those left after calibration; the values described in Sec.\,\ref{sec:systematics} (visualized in appendix~\ref{app:cate}) do not lead to significant bias, as expected. We note the residual systematics ($m=10^{-3}(1+z)^2$ and $\Delta z=10^{-3}(1+z)^2$) here can reach $\sim10^{-2}$ at $z\sim2$, which is significantly larger than the common requirements of a Stage-IV survey \cite{yao2023csst}. But the input $(1+z)^2$ term is to mimic the small \& faint galaxies at high-$z$, which we consider a conservative assumption.

Overall, the most dominant observational bias comes from lens magnification, which is at similar order as the modeling accuracy, judging from the biases and the reduced-$\chi^2$ values. Although building an emulator for the ratio is beyond the scope of this paper, this also motivates us to include a further test, keeping all the other systematics but removing only lens magnification, to see the improvement. The results are shown in the next section.

\subsection{Flat \texorpdfstring{$\Lambda$}{Lambda}CDM constraints}
\label{sec:lcdm}

\begin{figure}[tbp]
	\centering
	\includegraphics[width=\textwidth]{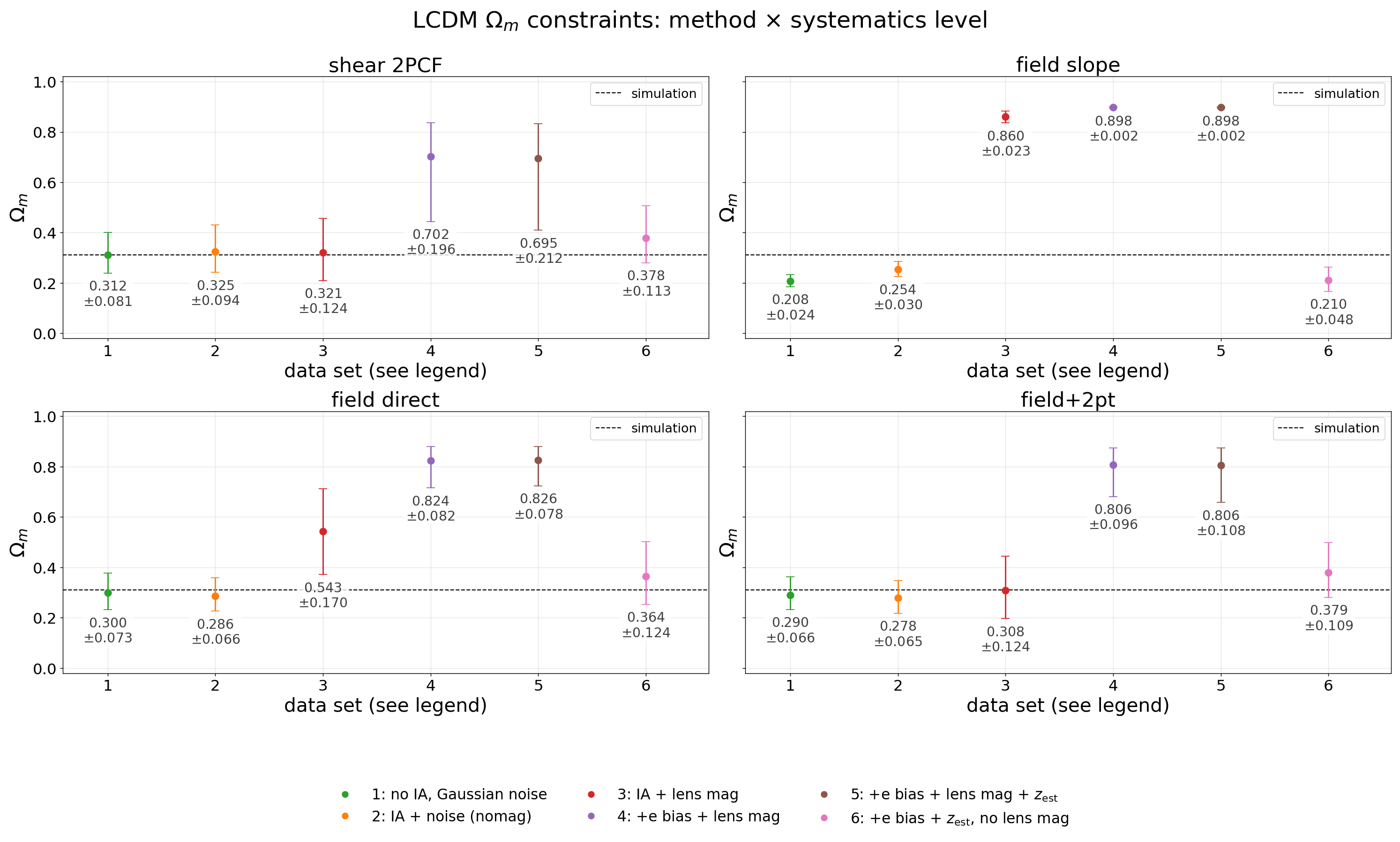}
	\caption{Flat $\Lambda$CDM constraints on $\Om$ for the four estimators (panels) and the six systematics levels (x-axis). The dashed horizontal line is the simulation truth.  The error bars are the 16th--84th percentile ranges of the marginal posteriors. We see the conventional shear 2PCF ratio is almost immune to IA, and slightly affected by residual shear and redshift biases, but the impact from the net contamination of lens magnification and residual shear bias is strong (top-left panel). Although the field-level slope ratio itself does not bias the ratio estimation (Fig.\,\ref{fig:compare}), its covariance is ill-shaped due to the linear assumption (appendix~\ref{app:cov}), leading to tight but strongly biased $\Omega_m$ constraints (top-right panel). By using the field-level direct ratio, the ill-shaped covariance due to the linear assumption is corrected, the field-level ratio reveals stronger constraining power than the 2PCF ratio when no systematics are added or only IA is included (bottom-left panel). But we note the contamination from lens magnification appears to have stronger impact than the 2PCF ratio. The field+two-point joint estimator collects the advantages from both sides, while it is still biased when both lens magnification and residual shear bias exist (bottom-right panel). The lens magnification is found to be the leading observational systematics, and for all cases the $\Omega_m$ is much less biased when it disappears.}
	\label{fig:om}
\end{figure}

We examine how the biases in Fig.\,\ref{fig:noia}, \ref{fig:ia}, \ref{fig:mag}, and \ref{fig:alleff} propagate to the cosmological constraints. Fig.\,\ref{fig:om} shows the flat $\Lambda$CDM constraints on $\Om$ for the
four estimators and the six systematics levels, with the sixth case representing the case in which only lens magnification is removed (for example, due to mitigation methods like proper weighting \cite{qin2024,qin2025}), while the others still exist. 

We first compare the 2PCF ratio (top-left panel) and the field-level direct ratio (bottom-left panel). In systematics level 1 (Gaussian shape noise only) and level 2 (IA added), the field-level direct ratio improves the statistical constraining power on $\Omega_m$ by $\sim10\%$ to $\sim40\%$ compared with the conventional 2PCF ratio, demonstrating the field-level information is important for the geometry studies with weak lensing. Especially when the non-Gaussian intrinsic shape is included (which we believe is more realistic than Gaussian shape noise), the shape noise contaminates the 2PCF ratio and the field-level ratio in different ways, which leads to the $\sim40\%$ improvement.

However, after including the lens magnification in the data (level 3), the field-level ratio is strongly contaminated, with the statistical error enlarged and a systematic error appearing. Meanwhile, the 2PCF ratio shows stronger stability against lens magnification, with the statistical error enlarged by $\sim32\%$ but no systematic error. Further tests with level 4 (residual shear bias) and level 5 (residual mean-$z$ bias) systematics leave $\Omega_m$ strongly over-estimated all the way towards the prior boundary, which suggests a total failure in the systematics mitigation and cosmological inference. Last but not least, in level 6 (keep all the other systematics but remove lens magnification), we find the results for both 2PCF ratio and the field-level ratio improved. Despite the fact that it is still slightly $\sim0.5\sigma$ biased with a larger errorbar than level 1 \& 2, it returns to an acceptable level. This further validates lens magnification as the most important observational systematic in lensing ratio studies.

By combining the field-level ratio and the 2PCF ratio, the resulting $\Omega_m$ constraints are better than any of them alone, in terms of both constraining power and bias. Overall, field-level information is not only useful for enhancing the cosmological output, but its comparison with conventional two-point statistics is also a useful tool for examining systematics.

One more thing to notice is the field-level slope ratio. In Fig.\,\ref{fig:compare} we showed the slope ratio gives almost identical ratio estimation as the direct ratio, and this is supported over a wide range of comparisons in Fig.\,\ref{fig:noia}, \ref{fig:ia}, \ref{fig:mag}, and \ref{fig:alleff}. But in terms of cosmological constraints, it is shown to be largely biased. This is due to the extra information introduced by the linear assumption. The linear assumption not only affects the single-value estimation, but also appears to have a strong impact on the covariance. In appendix~\ref{app:cov} we examine the covariance for all the ratio estimators, including the effect of truncating its smallest eigenmodes. Although they all have close values, the covariance of the slope ratio is strongly ill-shaped, with many nearly singular eigenvalues. This makes the inverse of the covariance very unstable, leading to tight but strongly biased constraints in Fig.\,\ref{fig:om}. This is why we choose the direct ratio without any prior assumption as the fiducial field-level ratio estimator. The triple-by-triple contribution to the cosmological shift is decomposed in appendix~\ref{app:decomp}.

% ===========================================================================
\section{Discussion}
\label{sec:discussion}
% ===========================================================================

We have compared different lensing ratio estimators, their sensitivity to observational systematics, and the resulting cosmological constraints. The goal is to provide a precursor exploration for Stage-IV surveys to constrain cosmic geometry.

\subsection{A new zero-lag field-level lensing ratio estimator}
\label{sec:newestimator}

We have developed a zero-lag field-level lensing ratio estimator that does not rely on machine learning, but is based only on the local $\kappa(\delta_g)$ relation (Fig.\,\ref{fig:field}). We have compared its performance with the conventional two-point ratio and possible variations of the field-level estimators. The comparison shows that converting the shear catalog into a convergence map loses a significant amount of information, mainly because of the finite pixel size. Starting from the same map, the field-level ratio extracts more information from the field than the map-level two-point ratio (Fig.\,\ref{fig:compare}). 

In terms of different field-level estimators, the slope ratio and the direct ratio are very close for a single ratio measurement (Fig.\,\ref{fig:compare} \& \ref{fig:noia}). At first sight this suggests that the linear galaxy bias assumption is not a problem. However, this assumption distorts the covariance (appendix~\ref{app:cov}) of the slope ratio and introduces an additional bias in the cosmological inference (Fig.\,\ref{fig:om}). The direct ratio is therefore the correct field-level choice.

When observational systematics are not significant, the field-level ratio improves the $\Om$ constraint by $10\%$--$40\%$ for different shape-noise cases (Fig.\,\ref{fig:om}), further emphasizing the importance of field-level information. We note that, although cosmic geometry itself does not contain the information from non-Gaussianity due to structure formation, the definition of lensing ratio can benefit from the extra non-Gaussian information in both the numerator and denominator, enhancing the ratio used to probe cosmic expansion. Although our zero-lag approximation does not contain the full field information, this improvement is still promising.

\subsection{Sensitivity to systematics}
\label{sec:sensitivity}

Even when no observational systematics are included, the current theoretical modeling of the ratio (either $\Rmez$ or $\Rgeo$) is biased (Fig.\,\ref{fig:noia}). In our test with an early Stage-IV-level mock catalog, this mismatch does not significantly bias the cosmological constraints (Fig.\,\ref{fig:om}), but it significantly increases the reduced-$\chi^2$ and may waste the strong constraining power of the field-level ratio 
\footnote{A quick estimation is, the reduced-$\chi^2$ is $\sim5$ in Fig.\,\ref{fig:noia}, while a good model should fit at $\sim1$, so if this is fully due to modeling bias, a good model might improve the statistical error in Fig.\,\ref{fig:om} by $\sim\sqrt{5}$.}. Accurate theoretical modeling
is therefore important. An emulator built from a large set of high-accuracy simulation mocks could substantially alleviate this problem~\cite{chen2025emulator1,chen2025emulator2,zhou2025emulator3}, and provide a basis for more precise geometry constraints, such as $\Omega_k$, $w_0$, and $w_a$.

When observational systematics are present, their relative importance for the lensing ratio analysis is: lens magnification \(>\) residual shear bias \(\sim\) residual mean-$z$ bias \(>\) intrinsic alignment (Fig.\,\ref{fig:mag}, \ref{fig:alleff}, \& \ref{fig:om}). Lens magnification significantly increases the bias of the field-level ratio (Fig.\,\ref{fig:mag}), especially at high-$z$, because it modulates the lens counts that the field-level fit uses as a template. Residual shear bias and residual mean-$z$ bias do not significantly increase the $\Om$ bias, but they enlarge the statistical errors (Fig.\,\ref{fig:om}). The bias from intrinsic alignment is negligible (Fig.\,\ref{fig:ia}), since the tomographic bins separate most lenses from their sources and the alignment contribution cancels partly between the two source bins. If the lens-magnification systematic can be mitigated, the other systematics will not introduce significant bias in the lensing ratio for the early data release of Stage-IV surveys (Fig.\,\ref{fig:om}). The comparisons between the field-level lensing ratio and the 2PCF lensing ratio could also be used as a diagnostic for systematics contamination.

\subsection{Future applications}
\label{sec:future}

The present results are based on the statistical and systematic error budget expected for the early data release of a Stage-IV survey (Sec.\,\ref{sec:data} and Sec.\,\ref{sec:systematics}), and do not represent the full capability of the lensing ratio.  In about \(5\)--\(10\) years, the full Stage-IV survey will provide a higher galaxy number density and a larger sky area, and the lensing ratio measurement with CMB lensing will become possible. These improvements will significantly increase the constraints on cosmic geometry.

The lensing ratio is particularly important for $\Om$ because $\Om$ is degenerate with $S_8$ and $H_0$ \cite{louis2025act,garciaquintero2025}.  It therefore plays an important role in the interpretation of both large-scale structure and cosmic expansion, addressing possible tensions, and affects the constraints on $\Omega_k$, $w_0$, and $w_a$ \cite{prat2026lensing}.

We note that the zero-lag pixel statistic, if evaluated without smoothing, is equivalent to assuming that each pixel is independent and contains only Gaussian information. Combining it with a Gaussian smoothing kernel brings some extra field-level information, because smoothing introduces correlations from the nearby field. However, to achieve a full field-level analysis, a (CNN- or transformer-like) multi-kernel study that varies with redshift and density environment is the key to maximizing the information. This requires an algorithm that retains the full pixel-field correlation. We leave this complex topic for future explorations.

\acknowledgments
% TODO: funding agencies, computing resources, and personal acknowledgments.
We thank DeepSeek (\url{https://www.deepseek.com/}) for accurate code modification and language polishing. We thank the members of the Jiutian simulation \cite{han2025,tan2025} and AKRA \cite{shi2024,shi2025,shi2026} teams for discussions. This work made use of the HEALPix \cite{gorski2005}, TreeCorr \cite{jarvis2004}, CCL \cite{chisari2019}, Astropy \cite{astropy2022}, NumPy \cite{harris2020}, SciPy \cite{virtanen2020}, and Matplotlib \cite{hunter2007} software packages. Numerical computations were performed on the computing cluster of the host institution. The analysis code and the reduced data products used in this paper are available from the authors upon reasonable request.

This work is supported by the China Manned Space Program
with grant No. CMS-CSST-2025-A03.
JY acknowledges the support from NSFC (Grant No. 12573006) and the State Key Laboratory of Dark
Matter Physics (DMEP2615). 
HYS acknowledges the support from NSFC of China under grant 11973070, the Shanghai Committee of Science and Technology grant No. 19ZR1466600 and Key Research Program of Frontier Sciences, CAS, Grant No. ZDBS-LY-7013. 
PZ acknowledges the support of NSFC (12595310) and the National Key R\&D Program of China (2023YFA1607800, 2023YFA1607801).

\clearpage
\appendix
% ===========================================================================
% appendices
% ===========================================================================
\renewcommand{\thefigure}{A\arabic{figure}}
\setcounter{figure}{0}
\renewcommand{\thetable}{A\arabic{table}}
\setcounter{table}{0}

% ---------------------------------------------------------------------------
\section{Smoothing-scale stability of the direct ratio}
\label{app:smoothing}
% ---------------------------------------------------------------------------

\begin{figure}[H]
\centering
\includegraphics[width=0.9\textwidth]{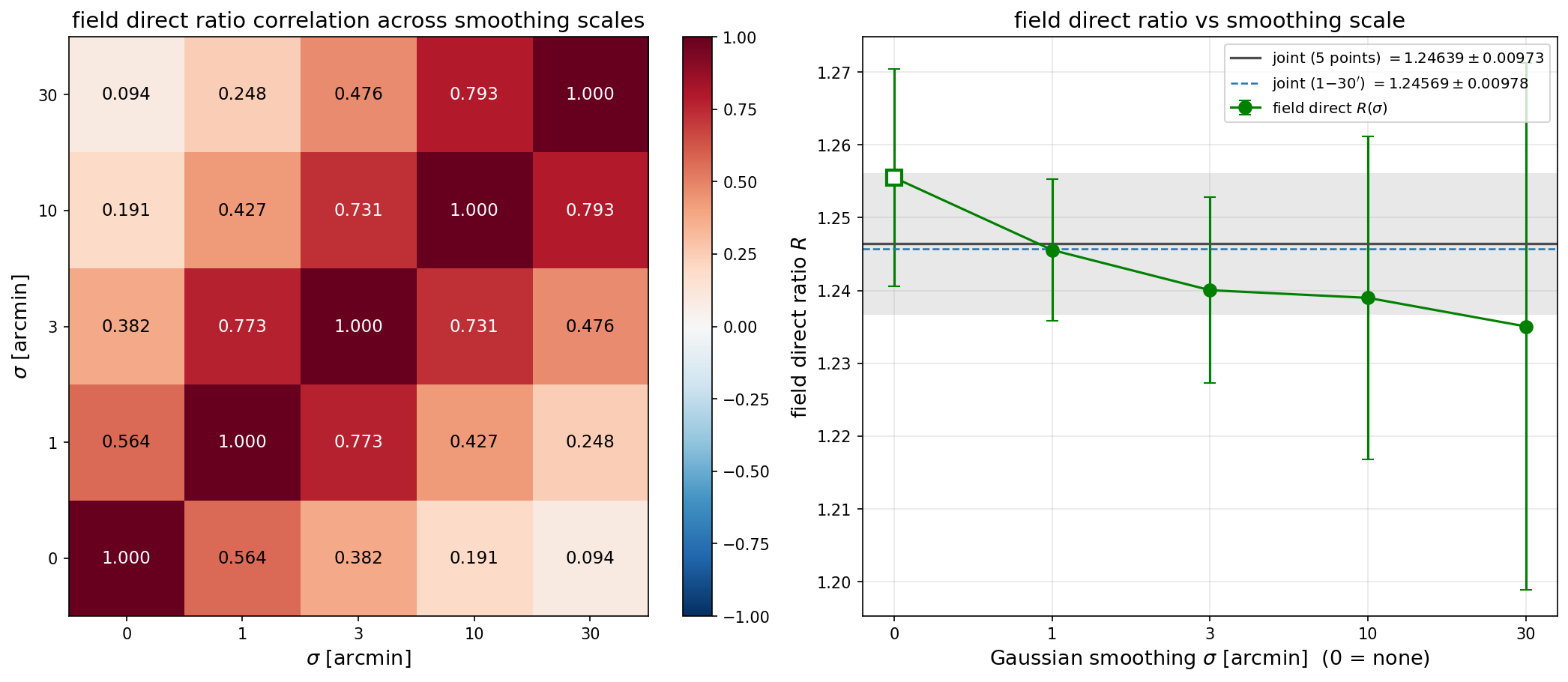}
\caption{Smoothing-scale stability of the direct ratio.  Left:
$5\times5$ correlation matrix of the direct ratio across the five smoothing scales ($0$ denotes no smoothing). Right: direct ratio as a function of the Gaussian smoothing scale; the shaded band is the joint GLS over the five scales, $\Rdirect=1.24639\pm0.00973$, and the dashed line is the joint GLS excluding the unsmoothed point. The measurement uses the IA+noise source maps (same as the previous Fig.\,\ref{fig:compare}), the unlensed lens positions, and the same 200-patch jackknife at every
scale. This result demonstrates that $1\arcmin$ smoothing is the optimal choice.}
\label{fig:smoothing}
\end{figure}

The standard direct ratio measurement uses the $\rm nside=4096$ pipeline with $1\arcmin$ Gaussian smoothing on the shear map. The smoothing scale determines how much the non-Gaussian field is converted into a Gaussian field and how strongly small scales are suppressed. To test the sensitivity to this choice, we repeat the direct estimator at $0$ (no smoothing), $1$, $3$, $10$, and $30\arcmin$, keeping the same IA+noise source maps, the unlensed (\texttt{nomag}) lens positions, and the same 200-patch jackknife at every scale (Fig.\,\ref{fig:smoothing}).  For the four smoothed points we use one reconstruction recipe: the same $1\arcmin$ reconstruction footprint, $\ell_{\rm max}=8192$, $\lambda=10^{-3}$, $\rm cgtol=2\times10^{-3}$, and a maximum of 150 iterations. The unsmoothed point is the raw reconstruction and necessarily follows a different pixelization; we keep it as a
cross-check rather than as part of the uniform scan.

The five scales are strongly correlated at neighboring smoothing lengths, with coefficients between $0.56$ and $0.79$; the unsmoothed map is only weakly correlated with the $30\arcmin$ map ($0.094$), and the full correlation matrix spans $0.094$--$0.793$. The individual values are
$\Rdirect=1.2555\pm0.0150$, $1.2456\pm0.0097$, $1.2400\pm0.0128$, $1.2390\pm0.0222$, and $1.2351\pm0.0362$ at $0$, $1$, $3$, $10$, and $30\arcmin$, respectively (Fig.\,\ref{fig:smoothing}). The joint GLS over the five scales gives $\Rdirect=1.24639\pm0.00973$.  Excluding the unsmoothed point gives $1.24569\pm0.00978$, a shift of only $0.06\%$. All five values are consistent with a constant, with the largest deviation at $30\arcmin$, about $0.9\%$ below the joint value. We therefore find no evidence for a smoothing-induced bias in the direct ratio over the range probed here.

The consistency between the fiducial $1\arcmin$ result and the joint result across different smoothing scales confirms that $1\arcmin$ is the optimal smoothing scale. On the other hand, this result illustrates how our estimator combines zero-lag correlations and Gaussian smoothing:
$\bullet$ compared with the fiducial $1\arcmin$ smoothing, the unsmoothed direct ratio in Fig.\,\ref{fig:smoothing} over-estimates the result, while the galaxy-convergence 2pt ratio in Fig.\,\ref{fig:compare} under-estimates the result, suggesting more (or at least different) information is included in the zero-lag correlation;
$\bullet$ the correlation coefficient between the $1\arcmin$-smoothing result and the unsmoothed result is relatively low, and the mean ratio differs by $\sim0.6\sigma$ in Fig.\,\ref{fig:smoothing}; this suggests that more information is included during the smoothing.

% ---------------------------------------------------------------------------
\section{Covariance structure and eigenmode truncation}
\label{app:cov}
% ---------------------------------------------------------------------------

\begin{figure}[t]
\centering
\includegraphics[width=\textwidth]{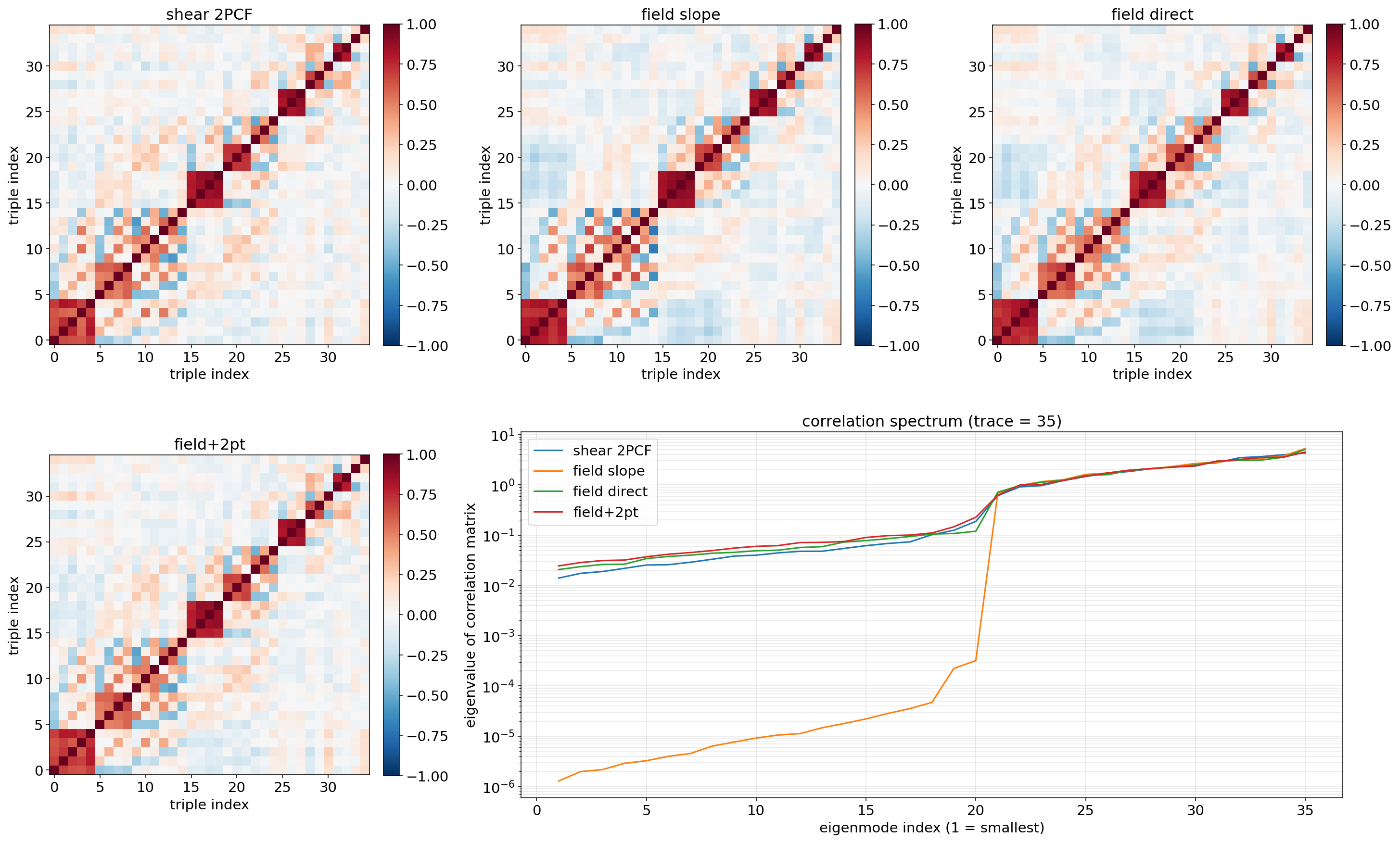}
\caption{Covariance structure of the 35-triple data vector for the full systematics sample (level 5, noise+IA+mag+e bias+$z_{\rm est}$). The four matrix panels show the diagonal-normalized $35\times35$ correlation matrices
for the 2PCF, the slope ratio, the direct ratio, and the joint estimator, which exhibit similar off-diagonal features. The bottom-right panel shows the eigenvalue spectra (trace $=35$). The slope ratio covariance has a significant drop after the 15 largest eigenvalues, which makes the inverse of the covariance strongly singular.}
\label{fig:covcorr}
\end{figure}

The covariance of the 35-triple data vector is central to the cosmological
interpretation, because the field-level estimators have a different
correlation structure from the two-point ratio.  Figure~\ref{fig:covcorr}
shows the diagonal-normalized $35\times35$ correlation matrices for the
2PCF, the slope ratio, the direct ratio, and the joint estimator, together
with their eigenvalue spectra; all matrices use the same 200-patch
partition.

The 2PCF covariance is well conditioned, with a condition number of about
$3\times10^2$, and its smallest mode carries less than $0.1\%$ of the
$\chi^2$ at the true cosmology.  The slope ratio covariance is very
different.  Its condition number is $5.3\times10^5$ in the clean sample and
about $4\times10^6$ in the catalog-$e$ sample shown in the figure, and its
smallest eigenmode carries $44\%$ of the $\chi^2$ at the true cosmology.  The
near-degenerate mode is not a numerical artifact.  It is the origin of the
artificially tight slope-only constraint discussed in
Sec.\,\ref{sec:sensitivity} and visible in Fig.\,\ref{fig:om}, and projecting
out only the smallest mode is not sufficient: a PCA with about ten
components is needed to bring the estimators into agreement.  The direct and
joint covariances are closer to the 2PCF case, which is one practical reason
why the direct ratio is adopted as the fiducial field-level estimator.

\begin{figure}[t]
\centering
\includegraphics[width=\textwidth]{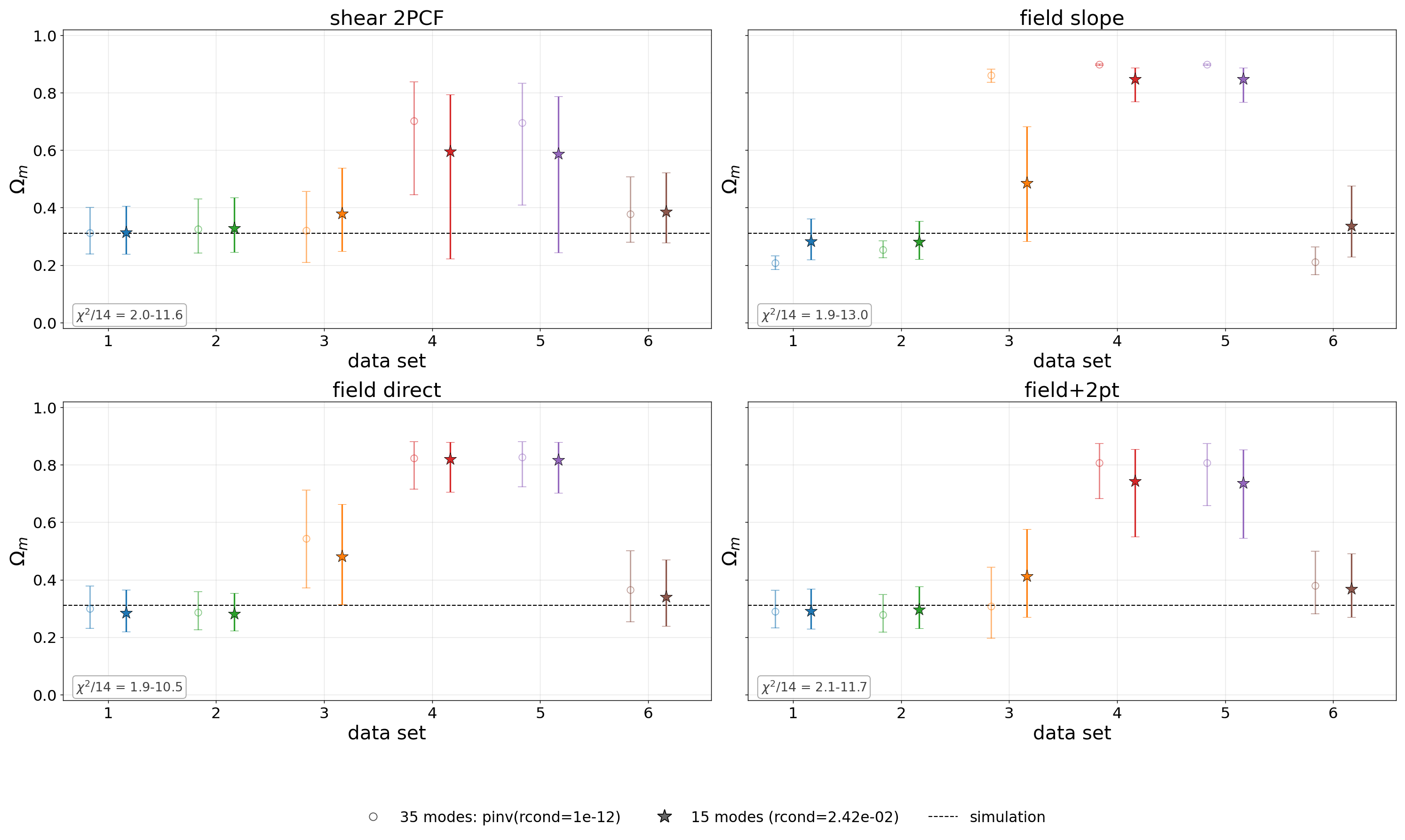}
\caption{Flat $\Lambda$CDM constraints on $\Om$ with the nominal covariance inverse (open circles; all 35 modes, $\rm rcond=10^{-12}$) and with a 15-mode truncated pseudo-inverse (stars; $\rm rcond=2.42\times10^{-2}$).  The panels show the four estimators and the six systematics levels of Fig.\,\ref{fig:om}. The truncation mainly affects the slope ratio, whose covariance contains a numerical null space; the 2PCF, direct, and joint
constraints are relatively stable.}
\label{fig:trunc}
\end{figure}

Figure~\ref{fig:covcorr} also shows that the slope-ratio spectrum drops rapidly at rank 15, while the other estimators have smooth spectra. This motivates a stability test of the covariance inversion: we keep only the first 15 eigenvalues and use a truncated pseudo-inverse ($\rm rcond=2.42\times10^{-2}$, with the Hartlap factor recomputed for $p=15$), and compare the resulting $\Om$ constraints with the nominal all-35-mode inversion (Fig.\,\ref{fig:trunc}). The field-slope results improve markedly, demonstrating that the problem in Fig.\,\ref{fig:om} is caused by the covariance. By contrast, the biases produced by lens magnification and shear bias are not significantly affected by the covariance truncation, indicating that they are not primarily covariance-driven. This leads to the following bias test.

% ---------------------------------------------------------------------------
\section{Diagnosing the large changes in $\Omega_m$}
\label{app:decomp}
% ---------------------------------------------------------------------------

Figure~\ref{fig:om} shows large changes in $\Om$ when lens magnification (level 3) and residual shear bias (level 4) are added. Appendix~\ref{app:cov} shows that these changes are not caused by the singular covariance: the 15-mode truncated pseudo-inverse improves the slope-ratio constraint, but it does not remove the shifts produced by lens magnification or shear bias. We therefore examine the origin of these shifts directly.

\begin{figure}[t]
\centering
\includegraphics[width=\textwidth]{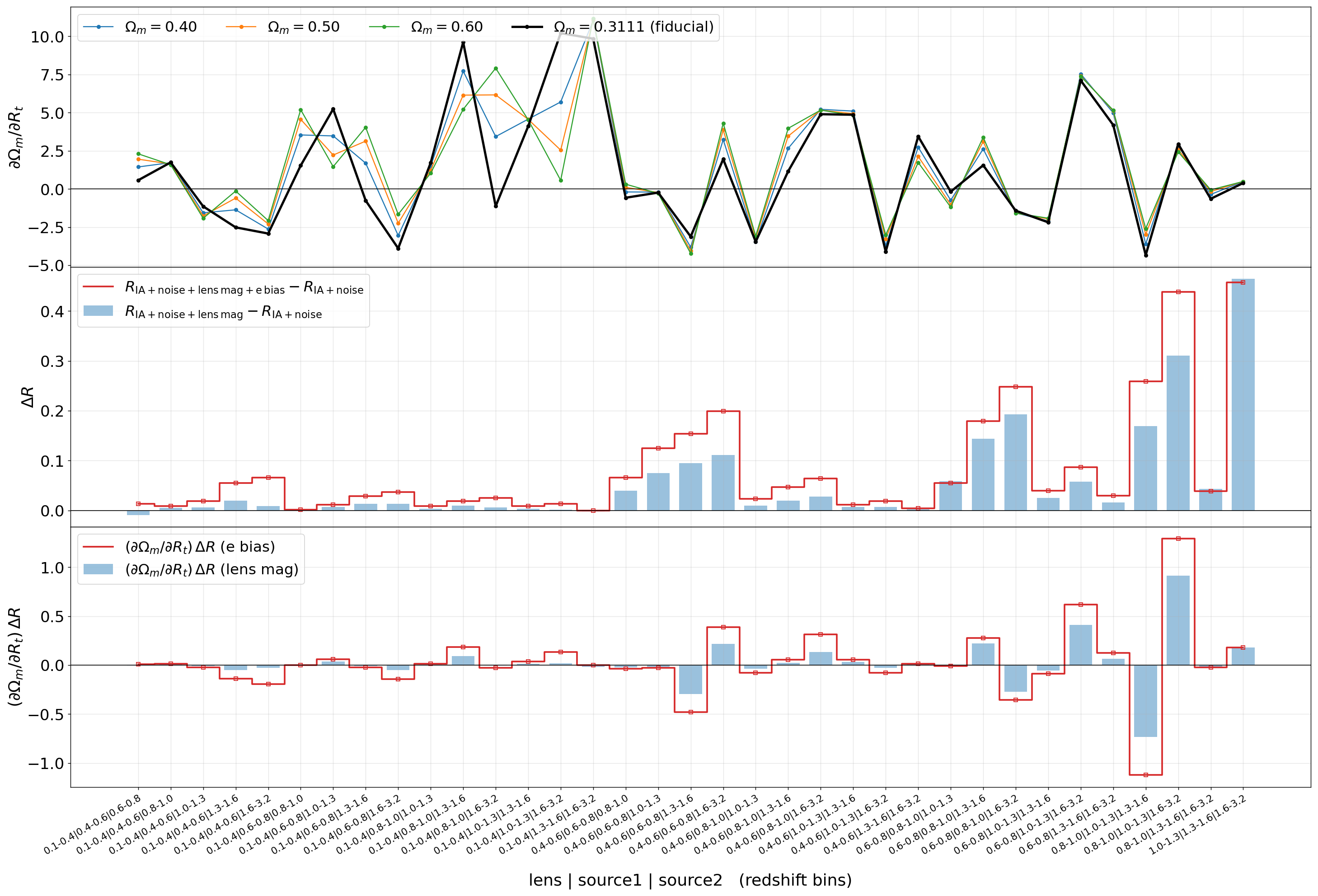}
\caption{Ratio-change decomposition for the field-direct estimator. Top: $\partial\Om/\partial R_t$ at $\Om=0.3111$, $0.40$, $0.50$, and $0.60$. Middle: ratio changes from lens magnification and from the catalog-$e$ bias relative to the IA+noise sample.  Bottom: the products $(\partial\Om/\partial R_t)\,\Delta R$, which show the contribution of each triple to the cosmological shift. The derivative and the ratio bias both vary strongly across triples, and their product can exceed unity.}
\label{fig:decomp}
\end{figure}

The top panel of Fig.\,\ref{fig:decomp} shows $\partial\Om/\partial R_t$ at several fiducial $\Om$ values. The derivative changes substantially with $\Om$ and across the 35 triples, and for some triples it changes by nearly an order of magnitude over the range shown. The mapping from a ratio change $\Delta R$ to a cosmological shift $\Delta\Om$ is therefore highly nonlinear, and $\Om$ is much more sensitive to some triples than to others.

The middle panel compares the field-direct ratio in systematics levels 2, 3, and 4, and shows that the ratio bias itself is also strongly triple-dependent. The bottom panel multiplies the derivative by this bias, which shows a simplified (ignoring the total covariance) first-order linear bias that is propagated to $\Om$. The per-triple contribution reaches about $0.5$ and in some cases exceeds $1$, so the ratio bias is already large enough to produce a significant change in $\Om$, while such a large bias activates the complicated non-linear $\Om-R$ response and cannot be directly visualized.

\begin{figure}[t]
\centering
\includegraphics[width=0.78\textwidth]{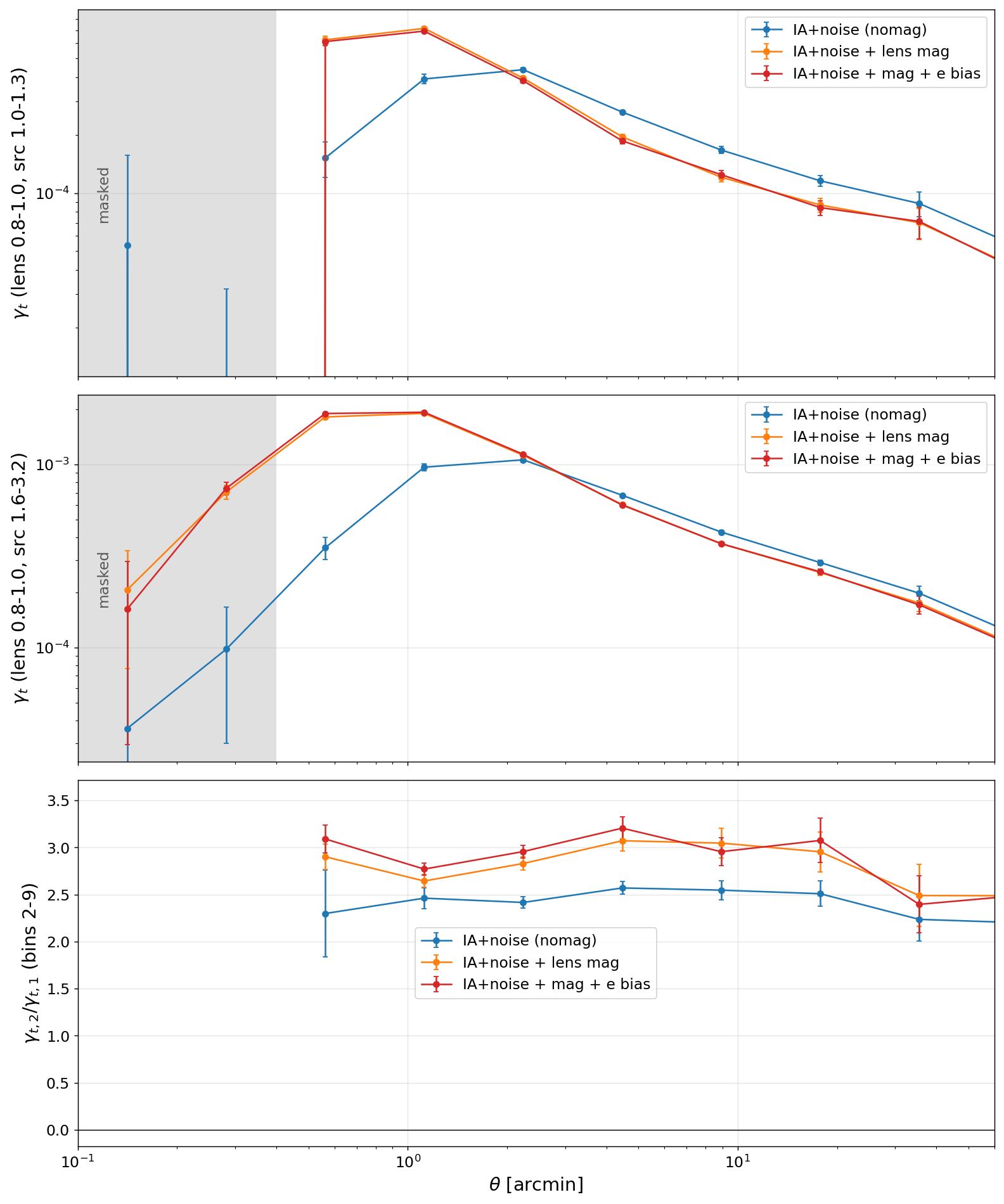}
\caption{Tangential shear for the lens bin $0.8$--$1.0$ and source bins $1.0$--$1.3$ (top) and $1.6$--$3.2$ (middle), and their ratio in bins 2--9 (bottom). Three data sets are shown: IA+noise without lens magnification (systematics level 2), with lens magnification (level 3), and with lens magnification plus the shear bias (level 4).  The shaded region at $\theta<0.4\arcmin$ is excluded from the ratio.}
\label{fig:gammat}
\end{figure}

To understand the largest ratio change, we select the triple
$0.8$--$1.0\,|\,1.0$--$1.3\,|\,1.6$--$3.2$ (the third-to-last triple in the bottom panel of Fig.\,\ref{fig:decomp}) and show its tangential shear 2pt measurements in Fig.\,\ref{fig:gammat}. Lens magnification changes both the amplitude and the slope of the two correlation functions, and the changes in the numerator and denominator are not equal. As a result, both the overall amplitude and the scale dependence of their ratio change significantly. This is the
two-point origin of the large $\Delta R$ that propagates into the cosmological constraint.

% ---------------------------------------------------------------------------
\section{Residual shear and mean-\texorpdfstring{$z$}{z} bias model}
\label{app:cate}
% ---------------------------------------------------------------------------

\begin{figure}[t]
	\centering
	\includegraphics[width=\textwidth]{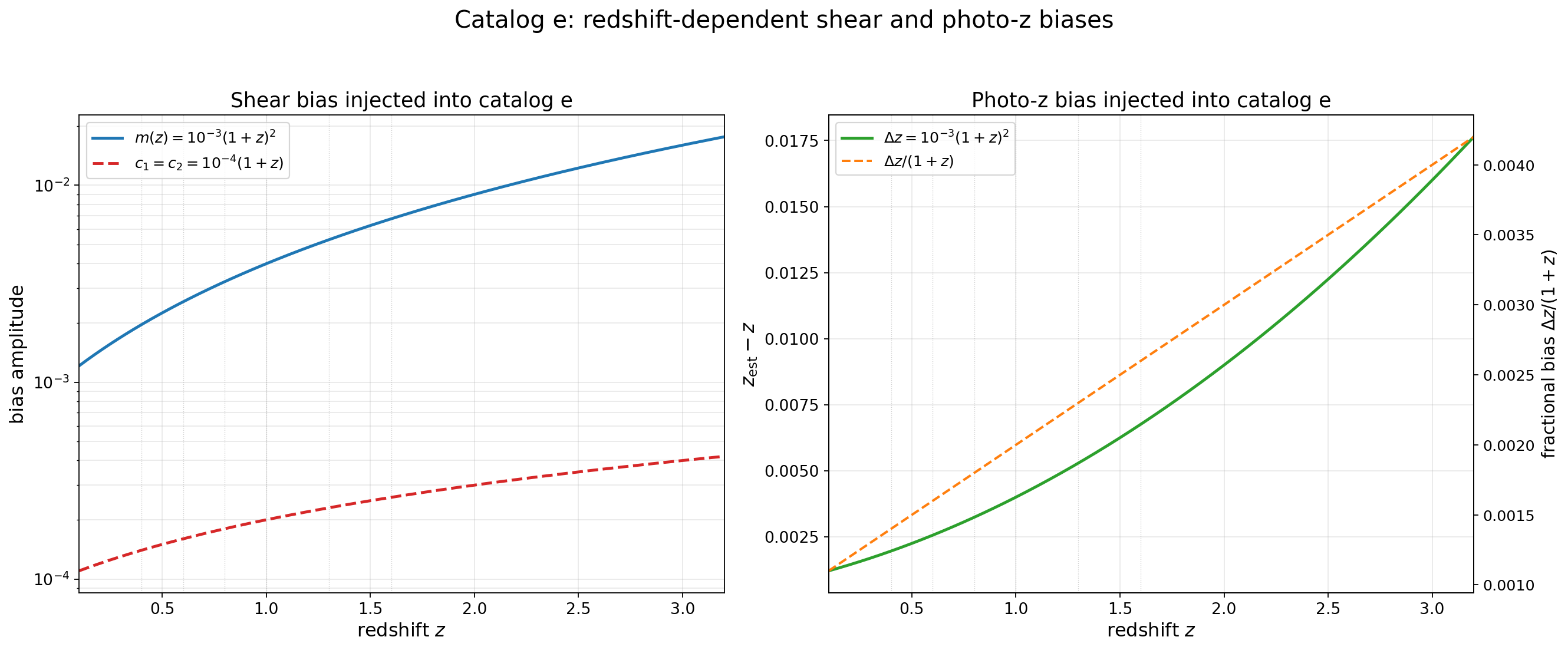}
	\caption{Redshift dependence of the source-side shear bias (level 4) and mean-$z$ bias (level 5) injected into the samples. Left: $m(z)$ and $c_1=c_2$. Right: $z_{\rm est}-z$ (solid) and $\Delta z/(1+z)$ (dashed). These biases approximately represent the mean requirements for systematics control, while the redshift dependence is intended to mimic the worse observational conditions at high-$z$, where galaxies have lower flux, smaller size, and fewer spec-$z$ matches.}
	\label{fig:cate}
\end{figure}

The shear bias (systematics level 4) and the mean-$z$ bias (level 5) are generated by the source-side biases described in Sec.\,\ref{sec:systematics}.  Figure~\ref{fig:cate} shows the redshift dependence of the individual terms used in those levels. The multiplicative bias $m(z)=10^{-3}(1+z)^2$ grows from $1.1\times10^{-3}$ at $z=0.1$ to $1.7\times10^{-2}$ at $z=3.2$, and the additive bias $c_1=c_2=10^{-4}(1+z)$ grows from $1.1\times10^{-4}$ to $4.2\times10^{-4}$. The photo-$z$ bias $z_{\rm est}-z=10^{-3}(1+z)^2$ corresponds to a fractional bias $\Delta z/(1+z)=10^{-3}(1+z)$, about $0.1\%$ at $z=0.1$ and $0.4\%$ at $z=3.2$.

These curves are a model illustration rather than a measurement. They are chosen to represent the residual biases left after calibration and to test the stability of the estimators when the biases grow towards high-$z$. The
actual amplitude of the shear and photo-$z$ biases in a real survey is set by the image simulations and by the shear and redshift calibration; the adopted redshift dependence is therefore a conservative test of the algorithm rather than a prediction for a specific survey.

\bibliographystyle{unsrtnat}
\bibliography{refs}

@article{planck2018,
  author  = {{Planck Collaboration}},
  title   = {Planck 2018 results. {VI}. Cosmological parameters},
  journal = {Astron. Astrophys.},
  volume  = {641},
  pages   = {A6},
  year    = {2020},
  doi     = {10.1051/0004-6361/201833910},
  eprint  = {1807.06209},
  archivePrefix = {arXiv},
  primaryClass  = {astro-ph.CO}
}

@article{desy3,
  author  = {Abbott, T. M. C. and others},
  collaboration = {DES Collaboration},
  title   = {Dark Energy Survey Year 3 results: Cosmological constraints from galaxy clustering and weak lensing},
  journal = {Phys. Rev. D},
  volume  = {105},
  pages   = {023520},
  year    = {2022},
  doi     = {10.1103/PhysRevD.105.023520},
  eprint  = {2105.13549},
  archivePrefix = {arXiv},
  primaryClass  = {astro-ph.CO}
}

@article{kids1000,
  author  = {Asgari, Marika and others},
  collaboration = {KiDS Collaboration},
  title   = {{KiDS-1000} cosmology: Cosmic shear constraints and comparison between two point statistics},
  journal = {Astron. Astrophys.},
  volume  = {645},
  pages   = {A104},
  year    = {2021},
  doi     = {10.1051/0004-6361/202039070},
  eprint  = {2007.15633},
  archivePrefix = {arXiv},
  primaryClass  = {astro-ph.CO}
}

@article{hscy1,
  author  = {Hikage, Chiaki and others},
  collaboration = {HSC Collaboration},
  title   = {Cosmology from cosmic shear power spectra with Subaru Hyper Suprime-Cam first-year data},
  journal = {Publ. Astron. Soc. Jap.},
  volume  = {71},
  pages   = {43},
  year    = {2019},
  doi     = {10.1093/pasj/psz010},
  eprint  = {1809.09148},
  archivePrefix = {arXiv},
  primaryClass  = {astro-ph.CO}
}

@article{euclid2020,
  author  = {Blanchard, A. and others},
  collaboration = {Euclid Collaboration},
  title   = {Euclid preparation: {VII}. Forecast validation for Euclid cosmological probes},
  journal = {Astron. Astrophys.},
  volume  = {642},
  pages   = {A191},
  year    = {2020},
  doi     = {10.1051/0004-6361/202038071},
  eprint  = {1910.09273},
  archivePrefix = {arXiv},
  primaryClass  = {astro-ph.CO}
}

@article{lsstdesc2018,
  author  = {{LSST Dark Energy Science Collaboration}},
  title   = {The {LSST} Dark Energy Science Collaboration ({DESC}) Science Requirements Document},
  journal = {arXiv e-prints},
  year    = {2018},
  eprint  = {1809.01669},
  archivePrefix = {arXiv},
  primaryClass  = {astro-ph.CO}
}

@article{gong2019,
  author  = {Gong, Yan and others},
  title   = {Cosmology from the Chinese Space Station Optical Survey ({CSS-OS})},
  journal = {Astrophys. J.},
  volume  = {883},
  pages   = {203},
  year    = {2019},
  doi     = {10.3847/1538-4357/ab391e},
  eprint  = {1901.04634},
  archivePrefix = {arXiv},
  primaryClass  = {astro-ph.CO}
}

@article{zhan2011,
  author  = {Zhan, Hu},
  title   = {Consideration for a large-scale multi-color imaging and slitless spectroscopy survey on the Chinese space station and its application in dark energy research},
  journal = {Sci. Sin. Phys. Mech. Astron.},
  volume  = {41},
  pages   = {1441},
  year    = {2011},
  doi     = {10.1360/132011-961}
}

@article{bartelmann2001,
  author  = {Bartelmann, Matthias and Schneider, Peter},
  title   = {Weak gravitational lensing},
  journal = {Phys. Rept.},
  volume  = {340},
  pages   = {291},
  year    = {2001},
  doi     = {10.1016/S0370-1573(00)00082-X},
  eprint  = {astro-ph/9912508},
  archivePrefix = {arXiv}
}

@article{kilbinger2015,
  author  = {Kilbinger, Martin},
  title   = {Cosmology with cosmic shear observations: a review},
  journal = {Rept. Prog. Phys.},
  volume  = {78},
  pages   = {086901},
  year    = {2015},
  doi     = {10.1088/0034-4885/78/8/086901},
  eprint  = {1411.0115},
  archivePrefix = {arXiv},
  primaryClass  = {astro-ph.CO}
}

@article{jaintaylor2003,
  author  = {Jain, Bhuvnesh and Taylor, Andy},
  title   = {Cross-correlation tomography: measuring dark energy evolution with weak lensing},
  journal = {Phys. Rev. Lett.},
  volume  = {91},
  pages   = {141302},
  year    = {2003},
  doi     = {10.1103/PhysRevLett.91.141302},
  eprint  = {astro-ph/0306046},
  archivePrefix = {arXiv}
}

@article{bernsteinjain2004,
  author  = {Bernstein, Gary M. and Jain, Bhuvnesh},
  title   = {Dark energy constraints from weak lensing cross-correlation cosmography},
  journal = {Astrophys. J.},
  volume  = {600},
  pages   = {348},
  year    = {2004},
  doi     = {10.1086/379755},
  eprint  = {astro-ph/0309332},
  archivePrefix = {arXiv}
}

@article{zhang2005,
  author  = {Zhang, Jun and Hui, Lam and Stebbins, Albert},
  title   = {Isolating geometry in weak lensing measurements},
  journal = {Astrophys. J.},
  volume  = {635},
  pages   = {806},
  year    = {2005},
  doi     = {10.1086/497573},
  eprint  = {astro-ph/0312348},
  archivePrefix = {arXiv}
}

@article{sun2023,
  author  = {Sun, Zeyang and Zhang, Pengjie and Dong, Fuyu and Yao, Ji and Shan, Huanyuan and Jullo, Eric},
  title   = {An unbiased method of measuring the ratio of two data sets},
  journal = {Astrophys. J. Suppl.},
  volume  = {267},
  pages   = {21},
  year    = {2023},
  doi     = {10.3847/1538-4365/acda2a},
  eprint  = {2210.13717},
  archivePrefix = {arXiv},
  primaryClass  = {astro-ph.CO}
}

@article{shi2024,
  author  = {Shi, Yuan and Zhang, Pengjie and Sun, Zeyang and Wang, Yihe},
  title   = {Accurate kappa reconstruction algorithm for masked shear catalog},
  journal = {Phys. Rev. D},
  volume  = {109},
  pages   = {123530},
  year    = {2024},
  doi     = {10.1103/PhysRevD.109.123530}
}

@article{shi2025,
  author  = {Shi, Yuan and Zhang, Pengjie and Deng, Furen and Zhou, Shuren and Cai, Hongbo and Yao, Ji and Sun, Zeyang},
  title   = {{AKRA} 2.0: Accurate Kappa Reconstruction Algorithm for masked shear catalog},
  journal = {JCAP},
  volume  = {2025},
  number  = {07},
  pages   = {038},
  year    = {2025},
  doi     = {10.1088/1475-7516/2025/07/038}
}

@article{shi2026,
  author  = {Shi, Yuan and Zhang, Pengjie and Cui, Li and Qin, Jian and Yao, Ji},
  title   = {{AKRA} 3.0: A Matrix-Free Inversion Framework for Weak Lensing Mass Mapping and Its Application to {DES Y3} Data},
  journal = {arXiv e-prints},
  year    = {2026},
  eprint  = {2606.06175},
  archivePrefix = {arXiv},
  primaryClass  = {astro-ph.CO}
}

@article{kaiser1993,
  author  = {Kaiser, Nick and Squires, Gordon},
  title   = {Mapping the dark matter with weak gravitational lensing},
  journal = {Astrophys. J.},
  volume  = {404},
  pages   = {441},
  year    = {1993},
  doi     = {10.1086/172297}
}

@article{gorski2005,
  author  = {G{\'o}rski, K. M. and others},
  title   = {{HEALPix}: A Framework for High-Resolution Discretization and Fast Analysis of Data Distributed on the Sphere},
  journal = {Astrophys. J.},
  volume  = {622},
  pages   = {759},
  year    = {2005},
  doi     = {10.1086/427976},
  eprint  = {astro-ph/0409513},
  archivePrefix = {arXiv}
}

@article{jarvis2004,
  author  = {Jarvis, M. and Bernstein, G. and Jain, B.},
  title   = {The skewness of the aperture mass statistic},
  journal = {Mon. Not. Roy. Astron. Soc.},
  volume  = {352},
  pages   = {338},
  year    = {2004},
  doi     = {10.1111/j.1365-2966.2004.07926.x},
  eprint  = {astro-ph/0307393},
  archivePrefix = {arXiv}
}

@article{chisari2019,
  author  = {Chisari, Nora Elisa and others},
  title   = {Core Cosmology Library: Precision Cosmological Predictions for {LSST}},
  journal = {Astrophys. J. Suppl.},
  volume  = {242},
  pages   = {2},
  year    = {2019},
  doi     = {10.3847/1538-4365/ab1658},
  eprint  = {1812.05995},
  archivePrefix = {arXiv},
  primaryClass  = {astro-ph.CO}
}

@article{foremanmackey2013,
  author  = {Foreman-Mackey, Daniel and Hogg, David W. and Lang, Dustin and Goodman, Jonathan},
  title   = {emcee: The MCMC Hammer},
  journal = {Publ. Astron. Soc. Pac.},
  volume  = {125},
  pages   = {306},
  year    = {2013},
  doi     = {10.1086/670067},
  eprint  = {1202.3665},
  archivePrefix = {arXiv},
  primaryClass  = {astro-ph.IM}
}

@article{hartlap2007,
  author  = {Hartlap, J. and Simon, P. and Schneider, P.},
  title   = {Why your model parameter confidences might be too optimistic. Unbiased estimation of the inverse covariance matrix},
  journal = {Astron. Astrophys.},
  volume  = {464},
  pages   = {399},
  year    = {2007},
  doi     = {10.1051/0004-6361:20066170},
  eprint  = {astro-ph/0608064},
  archivePrefix = {arXiv}
}

@article{dodelson2013,
  author  = {Dodelson, Scott and Schneider, Michael D.},
  title   = {The effect of covariance estimator error on cosmological parameter constraints},
  journal = {Phys. Rev. D},
  volume  = {88},
  pages   = {063537},
  year    = {2013},
  doi     = {10.1103/PhysRevD.88.063537},
  eprint  = {1304.2593},
  archivePrefix = {arXiv},
  primaryClass  = {astro-ph.CO}
}

@article{sellentin2016,
  author  = {Sellentin, Elena and Heavens, Alan F.},
  title   = {Parameter inference with estimated covariance matrices},
  journal = {Mon. Not. Roy. Astron. Soc.},
  volume  = {456},
  pages   = {L132},
  year    = {2016},
  doi     = {10.1093/mnrasl/slv190},
  eprint  = {1511.05969},
  archivePrefix = {arXiv},
  primaryClass  = {astro-ph.CO}
}

@article{friedrich2018,
  author  = {Friedrich, Oliver and Eifler, Tim},
  title   = {Precision matrix expansion for covariance matrix estimation in cosmological analyses},
  journal = {Mon. Not. Roy. Astron. Soc.},
  volume  = {473},
  pages   = {4150},
  year    = {2018},
  doi     = {10.1093/mnras/stx2600},
  eprint  = {1708.05810},
  archivePrefix = {arXiv},
  primaryClass  = {astro-ph.CO}
}

@article{hirata2004,
  author  = {Hirata, Christopher M. and Seljak, Uro{\v s}},
  title   = {Intrinsic alignment-lensing interference as a contaminant of cosmic shear},
  journal = {Phys. Rev. D},
  volume  = {70},
  pages   = {063526},
  year    = {2004},
  doi     = {10.1103/PhysRevD.70.063526},
  eprint  = {astro-ph/0406275},
  archivePrefix = {arXiv}
}

@article{bridle2007,
  author  = {Bridle, Sarah and King, Lindsay},
  title   = {Dark energy constraints from cosmic shear power spectra: impact of intrinsic alignments on photometric redshift requirements},
  journal = {New J. Phys.},
  volume  = {9},
  pages   = {444},
  year    = {2007},
  doi     = {10.1088/1367-2630/9/12/444}
}

@article{heymans2006,
  author  = {Heymans, Catherine and others},
  title   = {The Shear Testing Programme -- I. Weak lensing analysis of simulated ground-based observations},
  journal = {Mon. Not. Roy. Astron. Soc.},
  volume  = {368},
  pages   = {1323},
  year    = {2006},
  doi     = {10.1111/j.1365-2966.2006.10198.x},
  eprint  = {astro-ph/0506112},
  archivePrefix = {arXiv}
}

@article{zuntz2018,
  author  = {Zuntz, J. and others},
  collaboration = {DES Collaboration},
  title   = {Dark Energy Survey Year 1 results: weak lensing shape catalogues},
  journal = {Mon. Not. Roy. Astron. Soc.},
  volume  = {481},
  pages   = {1149},
  year    = {2018},
  doi     = {10.1093/mnras/sty2211},
  eprint  = {1708.01533},
  archivePrefix = {arXiv},
  primaryClass  = {astro-ph.CO}
}

@article{hoyle2018,
  author  = {Hoyle, B. and others},
  collaboration = {DES Collaboration},
  title   = {Dark Energy Survey Year 1 results: redshift distributions of the weak-lensing source galaxies},
  journal = {Mon. Not. Roy. Astron. Soc.},
  volume  = {478},
  pages   = {592},
  year    = {2018},
  doi     = {10.1093/mnras/sty957},
  eprint  = {1708.01532},
  archivePrefix = {arXiv},
  primaryClass  = {astro-ph.CO}
}

@article{elvinpoole2023,
  author  = {Elvin-Poole, J. and others},
  collaboration = {DES Collaboration},
  title   = {Dark Energy Survey Year 3 results: Magnification modeling and impact on cosmological constraints from galaxy clustering and galaxy-galaxy lensing},
  journal = {Mon. Not. Roy. Astron. Soc.},
  volume  = {523},
  year    = {2023},
  eprint  = {2209.09782},
  archivePrefix = {arXiv},
  primaryClass  = {astro-ph.CO}
}

@article{lange2024,
  author  = {Lange, J. U. and others},
  title   = {Systematic Effects in Galaxy-Galaxy Lensing with {DESI}},
  journal = {arXiv e-prints},
  year    = {2024},
  eprint  = {2404.09397},
  archivePrefix = {arXiv},
  primaryClass  = {astro-ph.CO}
}

@article{porqueres2023,
  author  = {Porqueres, Natalia and Heavens, Alan and Mortlock, Daniel and Lavaux, Guilhem and Makinen, T. Lucas},
  title   = {Field-level inference of cosmic shear with intrinsic alignments and baryons},
  journal = {arXiv e-prints},
  year    = {2023},
  eprint  = {2304.04785},
  archivePrefix = {arXiv},
  primaryClass  = {astro-ph.CO}
}

@article{zhou2024,
  author  = {Zhou, Alan Junzhe and Li, Xiangchong and Dodelson, Scott and Mandelbaum, Rachel},
  title   = {Accurate field-level weak lensing inference for precision cosmology},
  journal = {arXiv e-prints},
  year    = {2023},
  eprint  = {2312.08934},
  archivePrefix = {arXiv},
  primaryClass  = {astro-ph.CO}
}

@article{mandelbaum2006,
  author  = {Mandelbaum, Rachel and Seljak, Uro{\v s} and Kauffmann, Guinevere and Hirata, Christopher M. and Brinkmann, Jonathan},
  title   = {Galaxy halo masses and their dependence on environment and morphology},
  journal = {Mon. Not. Roy. Astron. Soc.},
  volume  = {368},
  pages   = {715},
  year    = {2006},
  doi     = {10.1111/j.1365-2966.2006.10156.x},
  eprint  = {astro-ph/0511164},
  archivePrefix = {arXiv}
}

@article{prat2022,
  author  = {Prat, J. and others},
  collaboration = {DES Collaboration},
  title   = {Dark Energy Survey Year 3 results: Galaxy-galaxy lensing},
  journal = {Phys. Rev. D},
  volume  = {105},
  pages   = {083529},
  year    = {2022},
  doi     = {10.1103/PhysRevD.105.083529},
  eprint  = {2105.13546},
  archivePrefix = {arXiv},
  primaryClass  = {astro-ph.CO}
}

@article{han2025,
  author  = {Han, Jiaxin and others},
  title   = {The Jiutian simulations for the {CSST} extra-galactic surveys},
  journal = {Sci. China Phys. Mech. Astron.},
  volume  = {68},
  pages   = {09511},
  year    = {2025},
  eprint  = {2503.21368},
  archivePrefix = {arXiv},
  primaryClass  = {astro-ph.CO}
}

@article{qin2024,
  author  = {Qin, Jian and Zhang, Pengjie and Xu, Haojie and Yu, Yu and Yao, Ji and Ma, Ruijie and Shan, Huanyuan},
  title   = {Weak lensing reconstruction by counting {DECaLS} galaxies},
  journal = {Phys. Rev. D},
  volume  = {110},
  pages   = {103538},
  year    = {2024},
  eprint  = {2310.15053},
  archivePrefix = {arXiv},
  primaryClass  = {astro-ph.CO}
}

@article{qin2025,
  author  = {Qin, Jian and Zhang, Pengjie and Chen, Zhu and Fu, Liping and Yu, Yu and Xu, Haojie and Yao, Ji and Shi, Yuan},
  title   = {Meta-calibration of the cosmic magnification coefficient: toward unbiased weak lensing reconstruction by counting galaxies},
  journal = {arXiv e-prints},
  year    = {2025},
  eprint  = {2505.16420},
  archivePrefix = {arXiv},
  primaryClass  = {astro-ph.CO}
}

@article{chisari2025,
  author  = {Chisari, Nora Elisa},
  title   = {A rising tide: Intrinsic alignments since the turn of the millennium},
  journal = {Astron. Astrophys. Rev.},
  volume  = {33},
  pages   = {5},
  year    = {2025},
  eprint  = {2510.15738},
  archivePrefix = {arXiv},
  primaryClass  = {astro-ph.CO}
}

@article{newman2022,
  author  = {Newman, Jeffrey A. and Gruen, Daniel},
  title   = {Photometric Redshifts for Next-Generation Surveys},
  journal = {Ann. Rev. Astron. Astrophys.},
  volume  = {60},
  pages   = {363},
  year    = {2022},
  eprint  = {2206.13633},
  archivePrefix = {arXiv},
  primaryClass  = {astro-ph.IM}
}

@article{li2023skils,
  author  = {Li, Shun-Sheng and others},
  title   = {{KiDS-Legacy} calibration: unifying shear and redshift calibration with the {SKiLLS} multi-band image simulations},
  journal = {Astron. Astrophys.},
  volume  = {670},
  pages   = {A100},
  year    = {2023},
  eprint  = {2210.07163},
  archivePrefix = {arXiv},
  primaryClass  = {astro-ph.CO}
}

@article{liu2026baryon,
  author  = {Liu, R. Henry and others},
  title   = {Model-independent measurement of baryon gas fractions through galaxy-galaxy lensing and the kinematic {Sunyaev-Zel'dovich} effect},
  journal = {arXiv e-prints},
  year    = {2026},
  eprint  = {2609.22466},
  archivePrefix = {arXiv},
  primaryClass  = {astro-ph.CO}
}

@article{ding2025shear2cosmo,
  author  = {Ding, Jiacheng and Su, Chen and Yao, Ji and Zhang, Le and Shan, Huanyuan},
  title   = {The Shear-to-Cosmology Paradigm I. Hybrid Field-Level and Simulation-Based Framework for Weak Lensing Surveys},
  journal = {arXiv e-prints},
  year    = {2025},
  eprint  = {2511.22851},
  archivePrefix = {arXiv},
  primaryClass  = {astro-ph.CO}
}

@article{cao2026roman,
  author  = {Cao, Kaili and Weinberg, David H. and Miranda, Vivian and Dalal, Nihar and Eifler, Tim and Xu, Jiachuan and Bowden, Haley and Markovi{\v c}, Katarina},
  title   = {Fisher Forecasts for Cosmological Yields from $3\times2$pt Analysis of the Roman Space Telescope High Latitude Imaging Survey},
  journal = {arXiv e-prints},
  year    = {2026},
  eprint  = {2601.00438},
  archivePrefix = {arXiv},
  primaryClass  = {astro-ph.CO}
}

@article{wright2025,
  author  = {Wright, Angus H. and others},
  title   = {{KiDS-Legacy}: Cosmological constraints from cosmic shear with the complete Kilo-Degree Survey},
  journal = {Astron. Astrophys.},
  volume  = {703},
  pages   = {A158},
  year    = {2025},
  eprint  = {2503.19441},
  archivePrefix = {arXiv},
  primaryClass  = {astro-ph.CO}
}

@article{anbajagane2025,
  author  = {Anbajagane, D. and others},
  title   = {The {DECADE} cosmic shear project I: A new weak lensing shape catalog of 107 million galaxies},
  journal = {arXiv e-prints},
  year    = {2025},
  eprint  = {2502.17674},
  archivePrefix = {arXiv},
  primaryClass  = {astro-ph.CO}
}

@article{yao2023kids1000,
  author  = {Yao, Ji and Shan, Huanyuan and Zhang, Pengjie and Liu, Xiangkun and Heymans, Catherine and Joachimi, Benjamin and Asgari, Marika and Bilicki, Maciej and Hildebrandt, Hendrik and Kuijken, Konrad and Tr{\"o}ster, Tilman and van den Busch, Jan Luca and Wright, Angus},
  title   = {{KiDS-1000}: cross-correlation with {Planck} cosmic microwave background lensing and intrinsic alignment removal with self-calibration},
  journal = {Astron. Astrophys.},
  volume  = {673},
  pages   = {A111},
  year    = {2023},
  eprint  = {2301.13437},
  archivePrefix = {arXiv},
  primaryClass  = {astro-ph.CO}
}

@article{chen2026cluster,
  author  = {Chen, Xinyue and Li, Ran and Wang, Qiao and Lin, Hengjie and Gao, Liang and Liu, Xiangkun and Shan, Huanyuan and Yao, Ji and Li, Ming},
  title   = {Mapping Dark Matter in Merging Clusters with Weak Lensing for {CSST}: Morphology and Filament Recovery},
  journal = {arXiv e-prints},
  year    = {2026},
  eprint  = {2610.17857},
  archivePrefix = {arXiv},
  primaryClass  = {astro-ph.CO}
}

@article{scognamiglio2026,
  author  = {Scognamiglio, Diana and others},
  title   = {{3D} dark matter map of the {JWST COSMOS-Web} field using diffusion-based generative priors},
  journal = {arXiv e-prints},
  year    = {2026},
  eprint  = {2611.03786},
  archivePrefix = {arXiv},
  primaryClass  = {astro-ph.CO}
}

@article{xu2026neutrino,
  author  = {Xu, Wenshuo and Zhao, Cheng and Su, Chen and Shan, Huanyuan and Liu, Yu},
  title   = {Constraining Neutrino Mass with the Void Weak Lensing Effect},
  journal = {arXiv e-prints},
  year    = {2026},
  eprint  = {2603.09278},
  archivePrefix = {arXiv},
  primaryClass  = {astro-ph.CO}
}

@article{su2025void,
  author  = {Su, Chen and Shan, Huanyuan and Zhao, Cheng and Xu, Wenshuo and Zhang, Jiajun},
  title   = {Cosmological Constraints with Void Lensing I: the Simulation-Based Inference Framework},
  journal = {Astron. Astrophys.},
  volume  = {699},
  pages   = {A174},
  year    = {2025},
  eprint  = {2504.15149},
  archivePrefix = {arXiv},
  primaryClass  = {astro-ph.CO}
}

@article{mead2020,
  author  = {Mead, Alexander and Brieden, Samuel and Tr{\"o}ster, Tilman and Heymans, Catherine},
  title   = {{HMcode-2020}: Improved modelling of non-linear cosmological power spectra with baryonic feedback},
  journal = {Mon. Not. Roy. Astron. Soc.},
  volume  = {502},
  pages   = {1401},
  year    = {2021},
  doi     = {10.1093/mnras/stab082},
  eprint  = {2009.01858},
  archivePrefix = {arXiv},
  primaryClass  = {astro-ph.CO}
}

@article{zhou2023halo,
  author  = {Zhou, Yifeng and Han, Jiaxin},
  title   = {A physical and concise halo model based on the depletion radius},
  journal = {arXiv e-prints},
  year    = {2023},
  eprint  = {2303.10886},
  archivePrefix = {arXiv},
  primaryClass  = {astro-ph.CO}
}

@article{chen2025emulator2,
  author  = {Chen, Zhao and Yu, Yu},
  title   = {{CSST} Cosmological Emulator II: Generalized Accurate Halo Mass Function Emulation},
  journal = {Sci. China Phys. Mech. Astron.},
  volume  = {68},
  pages   = {109513},
  year    = {2025},
  eprint  = {2506.09688},
  archivePrefix = {arXiv},
  primaryClass  = {astro-ph.CO}
}

@article{chen2025emulator1,
  author  = {Chen, Zhao and Yu, Yu and Han, Jiaxin and Jing, Y. P.},
  title   = {{CSST} Cosmological Emulator I: Matter Power Spectrum Emulation with one percent accuracy to $k=10\,h/{\rm Mpc}$},
  journal = {Sci. China Phys. Mech. Astron.},
  volume  = {68},
  pages   = {289512},
  year    = {2025},
  eprint  = {2502.11160},
  archivePrefix = {arXiv},
  primaryClass  = {astro-ph.CO}
}

@article{zhou2025emulator3,
  author  = {Zhou, Shuren and Chen, Zhao and Yu, Yu},
  title   = {{CSST} Cosmological Emulator III: Hybrid {Lagrangian} Bias Expansion Emulation of Galaxy Clustering},
  journal = {arXiv e-prints},
  year    = {2025},
  eprint  = {2506.04671},
  archivePrefix = {arXiv},
  primaryClass  = {astro-ph.CO}
}

@article{wang2019clustering,
  author  = {Wang, Zhaoyu and Xu, Haojie and Yang, Xiaohu and Jing, Y. P. and Guo, Hong and Zheng, Zheng and Zu, Ying and Li, Zhigang and Liu, Chengze},
  title   = {Accurate Modeling of the Projected Galaxy Clustering in Photometric Surveys: I. Tests with Mock Catalogs},
  journal = {arXiv e-prints},
  year    = {2019},
  eprint  = {1906.00516},
  archivePrefix = {arXiv},
  primaryClass  = {astro-ph.CO}
}

@article{zheng2024feedback,
  author  = {Zheng, Yi and Zhang, Pengjie},
  title   = {Forecast constraints on the baryonic feedback effect from the future kinetic {Sunyaev-Zel'dovich} effect detection},
  journal = {arXiv e-prints},
  year    = {2024},
  eprint  = {2401.03510},
  archivePrefix = {arXiv},
  primaryClass  = {astro-ph.CO}
}

@article{zhang2026feedback,
  author  = {Zhang, Pengjie},
  title   = {Baryonic feedback suppression of the matter power spectrum: a three-parameter fitting formula and its single-parameter reduction},
  journal = {arXiv e-prints},
  year    = {2026},
  eprint  = {2609.00807},
  archivePrefix = {arXiv},
  primaryClass  = {astro-ph.CO}
}

@article{zhou2026dm,
  author  = {Zhou, Shuren and Zhang, Pengjie},
  title   = {Proposal to construct the dark-matter-only counterpart of the observed universe combining weak lensing and baryon censuses},
  journal = {Phys. Rev. Lett.},
  volume  = {137},
  pages   = {131002},
  year    = {2026},
  eprint  = {2511.10975},
  archivePrefix = {arXiv},
  primaryClass  = {astro-ph.CO}
}

@article{sanchez2021shear,
  author  = {S{\'a}nchez, C. and others},
  title   = {Dark Energy Survey Year 3 Results: Exploiting small-scale information with lensing shear ratios},
  journal = {arXiv e-prints},
  year    = {2021},
  eprint  = {2105.13542},
  archivePrefix = {arXiv},
  primaryClass  = {astro-ph.CO}
}

@article{schneider2016,
  author  = {Schneider, Peter},
  title   = {Generalized shear-ratio tests: A new relation between cosmological distances, and a diagnostic for a redshift-dependent multiplicative bias in shear measurements},
  journal = {Astron. Astrophys.},
  volume  = {592},
  pages   = {L6},
  year    = {2016},
  eprint  = {1603.04226},
  archivePrefix = {arXiv},
  primaryClass  = {astro-ph.CO}
}

@book{schneider1992,
  author    = {Schneider, Peter and Ehlers, J{\"u}rgen and Falco, Emilio E.},
  title     = {Gravitational Lenses},
  publisher = {Springer-Verlag},
  address   = {Berlin},
  year      = {1992}
}

@article{schneider2002,
  author  = {Schneider, Peter and van Waerbeke, Ludovic and Kilbinger, Martin and Mellier, Yannick},
  title   = {Analysis of two-point statistics of cosmic shear: I. Estimators and covariances},
  journal = {Astron. Astrophys.},
  volume  = {396},
  pages   = {1},
  year    = {2002},
  eprint  = {astro-ph/0206182},
  archivePrefix = {arXiv},
  primaryClass  = {astro-ph}
}

@article{tan2025,
  author  = {Tan, Zhenlin and Xie, Lizhi and Han, Jiaxin and Qiu, Yisheng and Fontanot, Fabio and De Lucia, Gabriella and Guo, Qi and Li, Qingyang and Zhou, Jiale and Jiang, Wenkang and Wang, Xin and He, Feihong and Jin, Chichuan and Jing, Yipeng and Li, Ming and Li, Xiaodong and Pei, Wenxiang and Wang, Wenting and Yang, Xiaohu and Yu, Yu},
  title   = {A semi-analytical mock galaxy catalog for the {CSST} extragalactic surveys from the {Jiutian} simulations},
  journal = {arXiv e-prints},
  year    = {2025},
  eprint  = {2511.03281},
  archivePrefix = {arXiv},
  primaryClass  = {astro-ph.CO}
}

@article{wei2018ia,
  author  = {Wei, Chengliang and Li, Guoliang and Kang, Xi and Luo, Yu and Xia, Qianli and Wang, Peng and Yang, Xiaohu and Wang, Huiyuan and Jing, Yipeng and Mo, Houjun and Lin, Weipeng and Wang, Yang and Li, Shijie and Lu, Yi and Zhang, Youcai and Lim, S. H. and Tweed, Dylan and Cui, Weiguang},
  title   = {Full-sky Ray-tracing Simulation of Weak Lensing Using {ELUCID} Simulations: Exploring Galaxy Intrinsic Alignment and Cosmic Shear Correlations},
  journal = {Astrophys. J.},
  volume  = {853},
  pages   = {25},
  year    = {2018},
  doi     = {10.3847/1538-4357/aaa40d},
  eprint  = {1801.03941},
  archivePrefix = {arXiv},
  primaryClass  = {astro-ph.CO}
}

@article{yao2020ia,
  author  = {Yao, Ji and Shan, Huanyuan and Zhang, Pengjie and Kneib, Jean-Paul and Jullo, Eric},
  title   = {Unveiling the Intrinsic Alignment of Galaxies with Self-Calibration and {DECaLS DR3} data},
  journal = {arXiv e-prints},
  year    = {2020},
  eprint  = {2002.09826},
  archivePrefix = {arXiv},
  primaryClass  = {astro-ph.CO}
}

@article{blazek2015,
  author  = {Blazek, Jonathan and Vlah, Zvonimir and Seljak, Uro{\v s}},
  title   = {Tidal alignment of galaxies},
  journal = {arXiv e-prints},
  year    = {2015},
  eprint  = {1504.02510},
  archivePrefix = {arXiv},
  primaryClass  = {astro-ph.CO}
}

@article{blazek2019,
  author  = {Blazek, Jonathan and MacCrann, Niall and Troxel, M. A. and Fang, Xiao},
  title   = {Beyond linear galaxy alignments},
  journal = {Phys. Rev. D},
  volume  = {100},
  pages   = {103506},
  year    = {2019},
  eprint  = {1708.09247},
  archivePrefix = {arXiv},
  primaryClass  = {astro-ph.CO}
}

@article{mccullough2026,
  author  = {McCullough, J. and others},
  title   = {Dark Energy Survey Year 3: Blue Shear},
  journal = {Phys. Rev. D},
  volume  = {113},
  year    = {2026},
  eprint  = {2410.22272},
  archivePrefix = {arXiv},
  primaryClass  = {astro-ph.CO}
}

@article{li2021noise,
  author  = {Li, HeKun and Zhang, Jun},
  title   = {Mitigating the Noise-Source Coupling Effect in Shear Measurement},
  journal = {Astrophys. J.},
  volume  = {911},
  pages   = {115},
  year    = {2021},
  doi     = {10.3847/1538-4357/abec6d},
  eprint  = {2012.10899},
  archivePrefix = {arXiv},
  primaryClass  = {astro-ph.CO}
}

@article{wright2025z,
  author  = {Wright, Angus H. and others},
  title   = {{KiDS-Legacy}: Redshift distributions and their calibration},
  journal = {Astron. Astrophys.},
  volume  = {703},
  pages   = {A144},
  year    = {2025},
  eprint  = {2503.19440},
  archivePrefix = {arXiv},
  primaryClass  = {astro-ph.CO}
}

@article{yao2023csst,
  author  = {Yao, Ji and Shan, Huanyuan and Li, Ran and Xu, Youhua and Fan, Dongwei and Liu, Dezi and Zhang, Pengjie and Yu, Yu and Wei, Chengliang and Hu, Bin and Li, Nan and Fan, Zuhui and Xu, Haojie and Guo, Wuzheng},
  title   = {{CSST} WL preparation I: forecast the impact from non-{Gaussian} covariances and requirements on systematics-control},
  journal = {arXiv e-prints},
  year    = {2023},
  eprint  = {2304.04489},
  archivePrefix = {arXiv},
  primaryClass  = {astro-ph.CO}
}

@article{fiedorowicz2022,
  author  = {Fiedorowicz, Pier and Rozo, Eduardo and Boruah, Supranta S. and Chang, Chihway and Gatti, Marco},
  title   = {{KaRMMa} -- kappa reconstruction for mass mapping},
  journal = {Mon. Not. Roy. Astron. Soc.},
  volume  = {512},
  pages   = {73},
  year    = {2022},
  doi     = {10.1093/mnras/stac468},
  eprint  = {2105.14699},
  archivePrefix = {arXiv},
  primaryClass  = {astro-ph.CO}
}

@article{jeffrey2021,
  author  = {Jeffrey, N. and others},
  title   = {Dark Energy Survey Year 3 results: Curved-sky weak lensing mass map reconstruction},
  journal = {Mon. Not. Roy. Astron. Soc.},
  volume  = {505},
  pages   = {4626},
  year    = {2021},
  doi     = {10.1093/mnras/stab1495},
  eprint  = {2105.13539},
  archivePrefix = {arXiv},
  primaryClass  = {astro-ph.CO}
}

@article{boruah2024,
  author  = {Boruah, Supranta S. and Fiedorowicz, Pier and Rozo, Eduardo},
  title   = {Bayesian mass mapping with weak lensing data using {KaRMMa}: validation with simulations and application to {Dark Energy Survey} Year 3 data},
  journal = {Phys. Rev. D},
  volume  = {110},
  pages   = {023524},
  year    = {2024},
  doi     = {10.1103/PhysRevD.110.023524},
  eprint  = {2403.05484},
  archivePrefix = {arXiv},
  primaryClass  = {astro-ph.CO}
}

@article{han2026miu,
  author  = {Han, W. G. and Zhao, An and Chen, Xinyue and Li, Ran and Li, Rui and Liu, Xiangkun and Chen, Zhao and Yu, Yu},
  title   = {{MIU2Net}: weak-lensing mass inversion using deep learning with nested {U}-structures},
  journal = {arXiv e-prints},
  year    = {2026},
  eprint  = {2601.13538},
  archivePrefix = {arXiv},
  primaryClass  = {astro-ph.CO}
}

@article{jeffrey2020deepmass,
  author  = {Jeffrey, Niall and Lanusse, Fran{\c c}ois and Lahav, Ofer and Starck, Jean-Luc},
  title   = {Deep learning dark matter map reconstructions from {DES SV} weak lensing data},
  journal = {Mon. Not. Roy. Astron. Soc.},
  volume  = {492},
  pages   = {5023},
  year    = {2020},
  doi     = {10.1093/mnras/staa127},
  eprint  = {1908.00543},
  archivePrefix = {arXiv},
  primaryClass  = {astro-ph.CO}
}

@article{white2026npe,
  author  = {White, Tim and Chandrashekaran, Shreyas and Avestruz, Camille and Regier, Jeffrey and {LSST Dark Energy Science Collaboration}},
  title   = {Neural Posterior Estimation for Tomographic Weak Lensing Mass Mapping},
  journal = {arXiv e-prints},
  year    = {2026},
  eprint  = {2609.07833},
  archivePrefix = {arXiv},
  primaryClass  = {astro-ph.CO}
}

@article{louis2025act,
  author  = {Louis, Thibaut and others},
  title   = {The Atacama Cosmology Telescope: {DR6} Power Spectra, Likelihoods and {$\Lambda$CDM} Parameters},
  journal = {arXiv e-prints},
  year    = {2025},
  eprint  = {2503.14452},
  archivePrefix = {arXiv},
  primaryClass  = {astro-ph.CO}
}

@article{garciaquintero2025,
  author  = {Garcia-Quintero, C. and others},
  title   = {Cosmological implications of {DESI DR2} {BAO} measurements in light of the latest {ACT DR6} {CMB} data},
  journal = {arXiv e-prints},
  year    = {2025},
  eprint  = {2504.18464},
  archivePrefix = {arXiv},
  primaryClass  = {astro-ph.CO}
}

@article{astropy2022,
  author  = {{Astropy Collaboration} and others},
  title   = {The {Astropy} Project: Sustaining and Growing a Community-oriented Open-source Project and the Latest Major Release (v5.0) of the Core Package},
  journal = {Astrophys. J.},
  volume  = {935},
  pages   = {167},
  year    = {2022},
  doi     = {10.3847/1538-4357/ac7c74},
  eprint  = {2206.14220},
  archivePrefix = {arXiv},
  primaryClass  = {astro-ph.IM}
}

@article{harris2020,
  author  = {Harris, Charles R. and others},
  title   = {Array programming with {NumPy}},
  journal = {Nature},
  volume  = {585},
  pages   = {357},
  year    = {2020},
  doi     = {10.1038/s41586-020-2649-2},
  eprint  = {2006.10256},
  archivePrefix = {arXiv},
  primaryClass  = {cs.MS}
}

@article{virtanen2020,
  author  = {Virtanen, Pauli and others},
  title   = {{SciPy} 1.0: fundamental algorithms for scientific computing in {Python}},
  journal = {Nature Methods},
  volume  = {17},
  pages   = {261},
  year    = {2020},
  doi     = {10.1038/s41592-019-0686-2},
  eprint  = {1907.10121},
  archivePrefix = {arXiv},
  primaryClass  = {cs.MS}
}

@article{hunter2007,
  author  = {Hunter, John D.},
  title   = {Matplotlib: A {2D} graphics environment},
  journal = {Comput. Sci. Eng.},
  volume  = {9},
  pages   = {90},
  year    = {2007},
  doi     = {10.1109/MCSE.2007.55}
}

@article{wandelt2004gibbs,
  author  = {Wandelt, Benjamin D. and Larson, David L. and Lakshminarayanan, Arun},
  title   = {Global, exact cosmic microwave background data analysis using {Gibbs} sampling},
  journal = {Phys. Rev. D},
  volume  = {70},
  pages   = {083511},
  year    = {2004},
  doi     = {10.1103/PhysRevD.70.083511},
  eprint  = {astro-ph/0310080},
  archivePrefix = {arXiv},
  primaryClass  = {astro-ph}
}

@article{millea2020delensing,
  author  = {Millea, Marius and Anderes, Ethan and Wandelt, Benjamin D.},
  title   = {Bayesian delensing delight: sampling-based inference of the primordial {CMB} and gravitational lensing},
  journal = {arXiv e-prints},
  year    = {2020},
  eprint  = {2002.00965},
  archivePrefix = {arXiv},
  primaryClass  = {astro-ph.CO}
}

@article{eriksen2004ilc,
  author  = {Eriksen, H. K. and Banday, A. J. and G{\'o}rski, K. M. and Lilje, P. B.},
  title   = {On foreground removal from the {Wilkinson Microwave Anisotropy Probe} data by an internal linear combination method: limitations and implications},
  journal = {Astrophys. J.},
  volume  = {612},
  pages   = {633},
  year    = {2004},
  doi     = {10.1086/422807},
  eprint  = {astro-ph/0403098},
  archivePrefix = {arXiv},
  primaryClass  = {astro-ph}
}

@article{cardoso2008smica,
  author  = {Cardoso, Jean-Fran{\c c}ois and Martin, Maude and Delabrouille, Jacques and Betoule, Marc and Patanchon, Guillaume},
  title   = {Component separation with flexible models---application to multichannel astrophysical observations},
  journal = {IEEE J. Sel. Topics Signal Process.},
  volume  = {2},
  pages   = {735},
  year    = {2008},
  doi     = {10.1109/JSTSP.2008.2005346},
  eprint  = {0803.1814},
  archivePrefix = {arXiv},
  primaryClass  = {astro-ph}
}

@article{white2016marked,
  author  = {White, Martin},
  title   = {A marked correlation function for constraining modified gravity models},
  journal = {J. Cosmol. Astropart. Phys.},
  volume  = {2016},
  number  = {11},
  pages   = {057},
  year    = {2016},
  doi     = {10.1088/1475-7516/2016/11/057},
  eprint  = {1609.08632},
  archivePrefix = {arXiv},
  primaryClass  = {astro-ph.CO}
}

@article{xiao2026marked,
  author  = {Xiao, Xu and Chen, Zhao and Yu, Yu and Li, Xiao-Dong and Zhang, Le},
  title   = {Improved Cosmological Constraints from Morphology-Based Marked Correlation Functions},
  journal = {arXiv e-prints},
  year    = {2026},
  eprint  = {2608.15083},
  archivePrefix = {arXiv},
  primaryClass  = {astro-ph.CO}
}

@article{qin2020copula,
  author  = {Qin, Jian and Yu, Yu and Zhang, Pengjie},
  title   = {The copula of the cosmological matter density field is non-{Gaussian}},
  journal = {Astrophys. J.},
  volume  = {897},
  pages   = {105},
  year    = {2020},
  doi     = {10.3847/1538-4357/ab952f},
  eprint  = {2006.06182},
  archivePrefix = {arXiv},
  primaryClass  = {astro-ph.CO}
}

@article{prat2026lensing,
  author  = {Prat, Judit and S{\'a}nchez, Carles},
  title   = {Lensing ratios as a geometric probe of dark energy in the {LSST} and {Simons Observatory} era},
  journal = {arXiv e-prints},
  year    = {2026},
  eprint  = {2611.32861},
  archivePrefix = {arXiv},
  primaryClass  = {astro-ph.CO}
}

\end{document}